\documentclass[twocolumn, tighten, floatfix]{aastex63}

\graphicspath{{./}{}}
\usepackage{amsmath}
\usepackage{CJK}
\usepackage{float}
\usepackage{lineno}
\usepackage{multirow}
\shorttitle{Nuclear Ring Star Formation}
\shortauthors{Song et al.}

\begin{document}

\title{A Comparison between Nuclear Ring Star Formation\\in LIRGs and Normal Galaxies with the Very Large Array}

\correspondingauthor{Yiqing Song}
\email{ys7jf@virginia.edu}

\begin{CJK*}{UTF8}{gbsn}
\author[0000-0002-3139-3041]{Y. Song}
\affiliation{Department of Astronomy, University of Virginia, 530 McCormick Road., Charlottesville, VA 22904, USA}

\author[0000-0002-1000-6081]{S. T. Linden}
\affiliation{Department of Astronomy, University of Massachusetts, LGRT-B 618, 710 North Pleasant Street, Amherst, MA 01003, USA}

\author[0000-0003-2638-1334]{A. S. Evans}
\affiliation{Department of Astronomy, University of Virginia, 530 McCormick Road., Charlottesville, VA 22904, USA}
\affiliation{National Radio Astronomy Observatory, 520 Edgemont Road, Charlottesville, VA 22903, USA }

\author[0000-0003-0057-8892]{L. Barcos-Mu\~{n}oz}
\affiliation{National Radio Astronomy Observatory, 520 Edgemont Road, Charlottesville, VA 22903, USA }

\author[0000-0003-3474-1125]{G. C. Privon}
\affiliation{National Radio Astronomy Observatory, 520 Edgemont Road, Charlottesville, VA 22903, USA }

\author[0000-0001-9163-0064]{I. Yoon}
\affiliation{National Radio Astronomy Observatory, 520 Edgemont Road, Charlottesville, VA 22903, USA }

\author[0000-0001-7089-7325]{E. J. Murphy}
\affiliation{National Radio Astronomy Observatory, 520 Edgemont Road, Charlottesville, VA 22903, USA }

\author[0000-0003-3917-6460]{K. L. Larson}
\affiliation{IPAC, Mail Code 314-6, Caltech, 1200 E. California Boulevard, Pasadena, CA
91125, USA}

\author[0000-0003-0699-6083]{T. D\'{i}az-Santos}
\affiliation{Institute of Astrophysics, Foundation for Research and Technology-Hellas, GR-71110, Heraklion, Greece}
\affiliation{N\'{u}cleo de Astronom\'{i}a de la Facultad de Ingenier\'{i}a y Ciencias, Universidad Diego Portales, Av. Ej\'{e}rcito Libertador 441, Santiago, Chile}
\affiliation{Chinese Academy of Sciences South America Center for Astronomy, National Astronomical Observatories, CAS, Beijing 100101, China}

\author[0000-0003-3498-2973]{L. Armus}
\affiliation{IPAC, Mail Code 314-6, Caltech, 1200 E. California Boulevard, Pasadena, CA
91125, USA}

\author[0000-0002-8204-8619]{Joseph M. Mazzarella}
\affiliation{IPAC, Mail Code 314-6, Caltech, 1200 E. California Boulevard, Pasadena, CA
91125, USA}

\author{J. Howell}
\affiliation{IPAC, Mail Code 314-6, Caltech, 1200 E. California Boulevard, Pasadena, CA
91125, USA}

\author[0000-0003-4268-0393]{H. Inami}
\affiliation{Hiroshima Astrophysical Science Center, Hiroshima University, 1-3-1 Kagamiyama, Higashi-Hiroshima, Hiroshima 739-8526, Japan}

\author[0000-0003-3638-8943]{N. Torres-Alb\`{a}}
\affiliation{Department of Physics \& Astronomy, Clemson University, 118 Kinard Laboratory, Clemson, SC 29634, USA}

\author[0000-0002-1912-0024]{V. U}
\affiliation{Department of Physics and Astronomy, University of California, Irvine, 4129 Frederick Reines Hall, Irvine, CA 92697, USA.}

\author[0000-0002-2688-1956]{V. Charmandaris}
\affiliation{Institute of Astrophysics, Foundation for Research and Technology-Hellas, GR-71110, Heraklion, Greece}
\affiliation{Department of Physics, University of Crete, Heraklion, 71003, Greece}

\author[0000-0003-3168-5922]{E. Momjian}
\affiliation{National Radio Astronomy Observatory, P.O. Box O, Socorro, NM 87801, USA}

\author[0000-0002-6149-8178]{J. McKinney}
\affiliation{Department of Astronomy, University of Massachusetts, LGRT-B 618, 710 North Pleasant Street, Amherst, MA 01003, USA}

\author[0000-0002-1568-579X]{D. Kunneriath}
\affiliation{National Radio Astronomy Observatory, 520 Edgemont Road, Charlottesville, VA 22903, USA }




\begin{abstract}
Nuclear rings are excellent laboratories for studying intense star
formation. We present results from a study of nuclear star-forming rings in five
nearby normal galaxies from the Star Formation in Radio Survey (SFRS) and four
local LIRGs from the Great Observatories All-sky LIRG Survey (GOALS) at sub-kpc
resolutions using VLA high-frequency radio continuum observations. We find that
nuclear ring star formation (NRSF) contributes 49 - 60\% of the total star
formation of the LIRGs, compared to 7 - 40\% for the normal galaxies. We
characterize a total of 58 individual star-forming regions in these rings, and
find that with measured sizes of 10 - 200\,pc, NRSF regions in the LIRGs have SFR and
$\Sigma_\mathrm{SFR}$ up to 1.7\,M$_\odot$yr$^{-1}$ and
402\,M$_\odot$yr$^{-1}$kpc$^{-2}$, respectively, which are about 10 times higher
than NRSF regions in the normal galaxies with similar sizes, and comparable to lensed
high-$z$ star-forming regions. At $\sim 100 - 300$\,pc
scales, we estimate low contributions ($< 50\%$) of thermal free-free emission
to total radio continuum emission at 33 GHz in the NRSF regions in the LIRGs,
but large variations possibly exist at smaller physical scales. Finally, using
archival sub-kpc resolution CO (J=1-0) data of nuclear rings in the normal
galaxies and NGC 7469 (LIRG), we find a large scatter in gas depletion times at similar
molecular gas surface densities, which tentatively points to a multi-modal star
formation relation on sub-kpc scales.
\end{abstract}



\section{Introduction}\label{sec:intro}
\end{CJK*}
At least one fifth of disk galaxies host star-forming nuclear rings
\citep{knapen05}. It is commonly accepted that nuclear rings result from a
non-axisymmetric gravitational potential in galaxy centers, which can be induced
by the presence of a stellar bar, strong spiral arms or tidal interaction
\citep[e.g.][]{combes85,shlosman90,athanassoula94,butacombes96,combes01}. Such
non-axisymmetry can drive large amounts of gas into the central region and
eventually develop a ring of dense gas and intense star formation (SF)
surrounding the galactic nucleus, likely at the location of the Inner Lindblad
Resonance \citep{kim12a,li15} or Nuclear Lindblad Resonance \citep{fukuda98}.
Optical and infrared (IR) studies of nuclear rings in nearby galaxies indicate
that they are prolific in producing young ($<$ 100\,Myr) and massive ($\sim
10^5\,M_\odot$) star clusters in episodic starbursts
\citep[e.g.][]{maoz96,buta2000,maoz01}. With large reservoirs of dense gas
present, nuclear rings are predicted to be long-lived despite the enhanced
massive SF \citep{garcia91,regan03,allard06,sarzi07}. Simulations have been used
to predict that even when subject to supernovae feedback, nuclear rings may
persist at Gyr timescales \citep{knapen95,seo13,seo14}. Therefore, nuclear rings
provide excellent opportunities to study extreme cases of SF in nearby systems.
\\
\indent As a result of long-lasting SF activity, nuclear rings can account for a
large fraction of the stellar mass and bolometric luminosity in their host
galaxies \citep[e.g.][]{garcia91,genzel95}. In the process of secular evolution,
nuclear ring SF (NRSF) often emerges as galactic pseudo-bulges slowly assemble
from disk material \citep{kormendy04}. Meanwhile, powerful NRSF has also been
seen in high-resolution observations of galaxy mergers
\citep[e.g.][]{genzel95,knapen04,haan13,herrero-illana14}, which is a
comparatively more dramatic evolutionary process. Simulations of galaxy mergers
have also proposed nuclear rings as a potential fueling mechanism of quasars
\citep{hopkins10}. Therefore formation of nuclear rings may represent a common and
critical phase in galaxy evolution, and properties of NRSF may provide insights
into the key dynamical processes associated with various evolutionary paths.\\
\indent While nuclear rings in nearby disk galaxies have been extensively
investigated in the optical and IR, studies of nuclear rings in galaxy
mergers are relatively scarce, making a consistent comparison of the two galaxy populations
difficult. Luminous and Ultra-luminous Infrared Galaxies (LIRGs:
$L_{\mathrm{IR}}(8-1000\mu\mathrm{m})> 10^{11}L_\odot$; ULIRGs:
$L_{\mathrm{IR}}(8-1000\mu\mathrm{m})> 10^{12}L_\odot$) in the local Universe,
which are often interacting or merging gas-rich spirals, have provided excellent
opportunities to study SF in mergers. However, heavy dust obscuration makes the
centers of these systems elusive at optical wavelengths \citep{sanders96}.
Meanwhile, measurements of the nuclei in U/LIRGs can still be heavily affected
by non-uniform dust extinction even in the IR
\citep{diaz-santos11,piqueras13,u19}.\\
\indent With the advancement of high-frequency radio interferometry, a highly
detailed, extinction-free view of the heavily obscured hearts of local LIRGs
becomes possible. In the present study, we make use of high-resolution ($\sim$
100\,pc) VLA observations to characterize and compare the SF properties of
nuclear rings hosted in five normal disk galaxies and four LIRGs in the local
Universe. Observations of these LIRGs are part of a new VLA campaign for
the Great Observatories All-sky LIRG Survey \citep[GOALS;][]{armus09} that
contains 68 local U/LIRGs, and the normal galaxies are observed with the Star
Formation in Radio Survey \citep[SFRS;][]{murphy12,murphy18,linden20}
of 56 nearby normal star-forming disk galaxies. These two projects together provide a
direct, high-resolution comparison of SF activity in interacting and isolated
galaxies in the local Universe. In this paper, we focus on comparing properties
of NRSF observed in these two surveys using high-frequency radio continuum as an
extinction-free tracer. A full study of the nuclear SF properties in the GOALS
VLA campaign will be presented in a forthcoming paper.\\
\indent This paper is divided into 7 sections. We present our sample selection,
observations and calibration procedures in Section \ref{sec:sample} and
\ref{sec:vladata}. In Section \ref{sec:analysis}, we describe the measurements
we perform to characterize the properties of the nuclear rings and the
individual NRSF regions, the results of which are presented in Section
\ref{sec:results}. In Section \ref{sec:discuss}, we discuss the limitations and
implications of the results. Finally, Section \ref{sec:summary} summarizes major
findings and conclusions. \\
\indent Throughout this work we adopt $H_0 = 70$\,km/s/Mpc,
$\Omega_\mathrm{matter} = 0.28$ and $\Omega_\Lambda = 0.72$ based on the
five-year WMAP result \citep{hinshaw09}. These parameters are used with the
3-attractor model \citep{mould00} to calculate the luminosity distances
and physical scales of the LIRGs in the sample.\\

\begin{figure*}[h!]
\figurenum{1(a)}
\centering
\epsscale{1.0}
\plotone{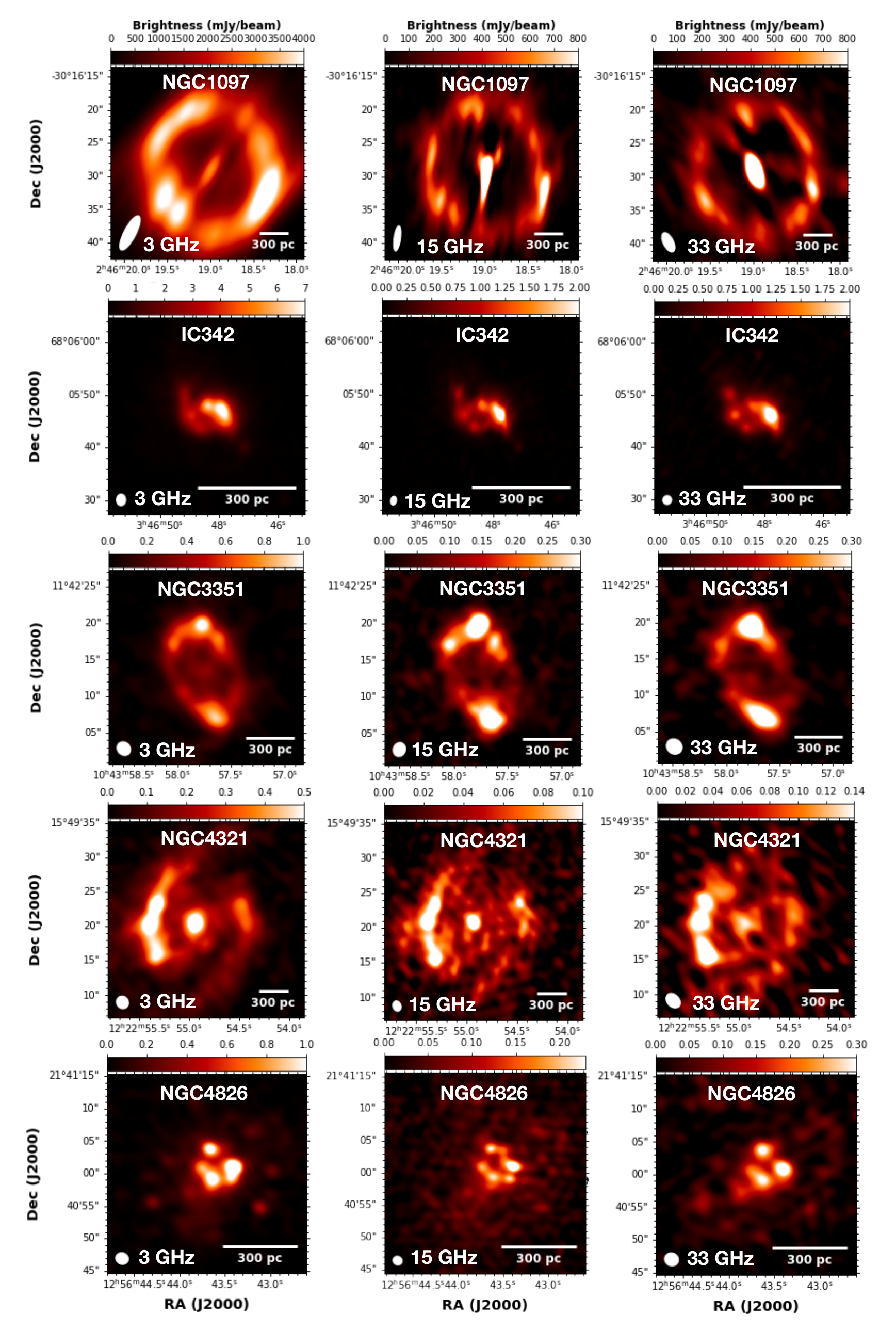}
\figcaption{Highest resolution Ka, Ku, and S band images of the five nuclear rings
hosted in the normal galaxies from SFRS. Each image is displayed with bilinear interpolation, in
units of mJy/beam, and the synthesized beam is represented by the white filled ellipse
on the lower left corner. The nuclear rings are well detected and resolved at
all three bands. \label{fig:natres1}}
\end{figure*}
      
      
\begin{figure*}[tbh!]
\figurenum{1(b)}
\centering
\epsscale{1.0}
\plotone{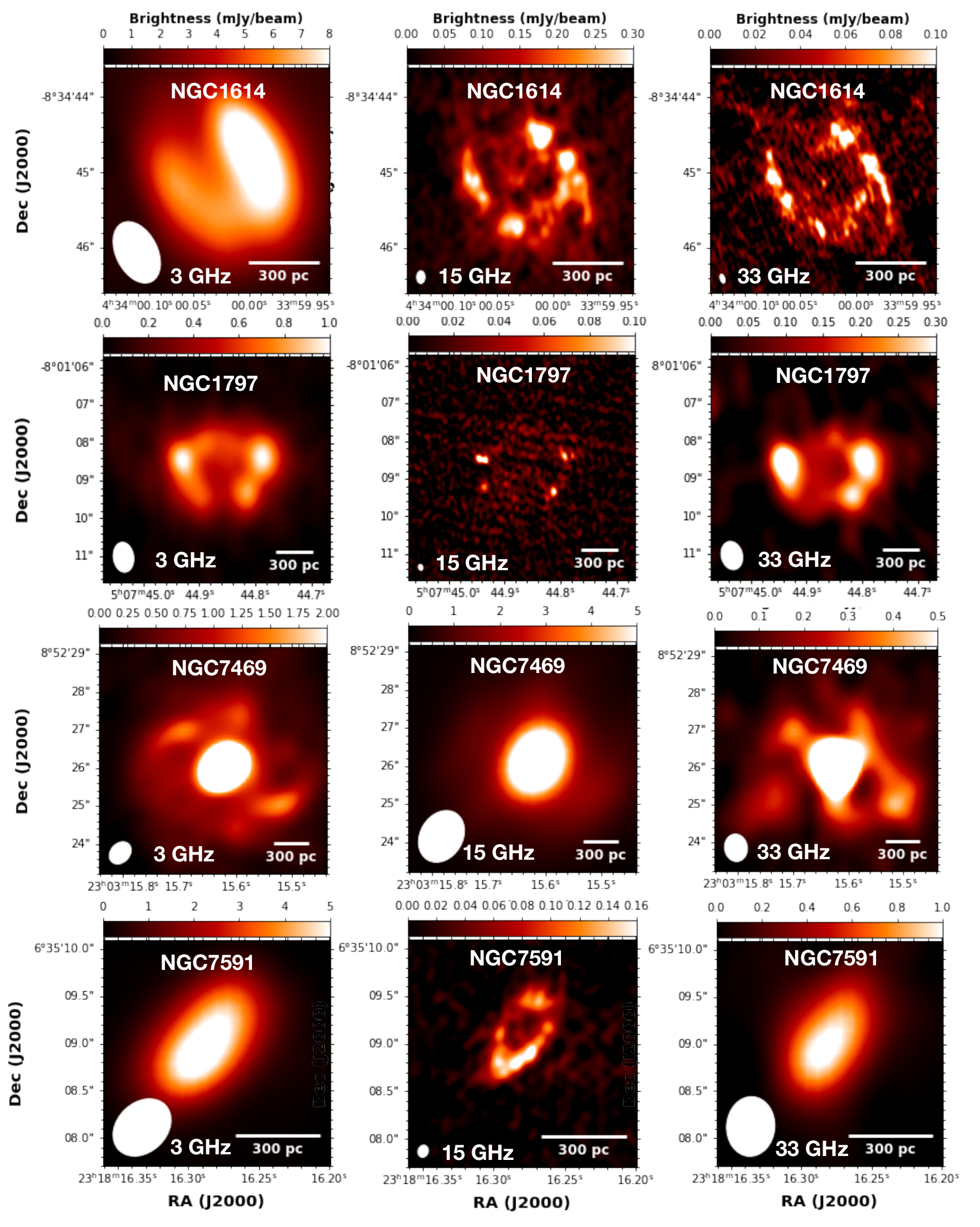}
\figcaption{Highest resolution Ka, Ku, and S band images for the four nuclear rings
hosted in LIRGs from GOALS. Each image is displayed with bilinear interpolation, in
units of mJy/beam, and the synthesized beam is represented by the white filled ellipse on the lower left
corner. All rings are detected and resolved at all three bands except for NGC
7591, whose nuclear ring was only resolved at Ku Band.\label{fig:natres2}}
\end{figure*}
\centerwidetable
\begin{deluxetable*}{lrrrrrccccl}[tbh!]
\tablecaption{Properties of Sample Galaxies \label{tab:sample}}
\tablecolumns{11}
\tablenum{1}
\tablewidth{0pt}
\tablehead{
\colhead{Host} &
\colhead{RA} &
\colhead{Dec} &
\colhead{$D_L$} &
\colhead{$\log_{10}(L_{\mathrm{IR}})$} &
\colhead{Scale} &
\colhead{Type} &
\colhead{$f^\mathrm{bol}_\mathrm{AGN}$} & 
\colhead{Survey} & 
\colhead{Morphology} &
\colhead{Merger}\\
\colhead{} &
\colhead{(J2000)} &
\colhead{(J2000)} &
\colhead{(Mpc)} &
\colhead{} &
\colhead{(pc/$\arcsec$)} &
\colhead{} &
\colhead{(\%)} & 
\colhead{} & 
\colhead{} &
\colhead{Stage}\\
\colhead{(1)} &
\colhead{(2)} &
\colhead{(3)} &
\colhead{(4)} &
\colhead{(5)} &
\colhead{(6)} &
\colhead{(7)} &
\colhead{(8)} & 
\colhead{(9)} &
\colhead{(10)} &
\colhead{(11)} 
}
\startdata 
NGC 1097 & 02h46m19.0s & -30d16m30s  & 14.2  & 10.56 & 69  & LINER/Sy1 & -         & SFRS   & SBb     & N$^*$\\
IC 342   & 03h46m48.5s & +68d05m47s  & 3.3   & 9.87  & 16  & HII       & -         & SFRS   & SABcd   & N\\
NGC 3351 & 10h43m57.7s &  +11d42m14s & 9.3   & 9.77  & 45  & HII       & -         & SFRS   & SBb     & N\\
NGC 4321 & 12h22m54.8s &  +15d49m19s & 14.3  & 10.33 & 70  & HII/LINER & -         & SFRS   & SABbc   & N\\
NGC 4826 & 12h56m43.6s &  +21d40m59s & 5.3   & 9.54  & 26   & LINER     & -         & SFRS   & SAab    & N\\   
NGC 1614 & 04h33m59.8s & -08d34m44s  & 67.8  & 11.65 & 323  & HII       & 12$\pm$8  & GOALS  & SBc pec & d\\
NGC 1797 & 05h07m44.9s & -08d01m09s  & 63.4  & 11.04 & 303  & HII       & \:\:6$\pm$3 & GOALS  & SBa pec & a\\
NGC 7469 & 23h03m15.6s & +08d52m26s  & 70.8  & 11.58 & 332  & Sy1       & 24$\pm$6  & GOALS  & SABa    & a\\ 
NGC 7591 & 23h18m16.3s & +06d35m09s  & 71.4  & 11.12 & 335  & LINER     & \:\:9$\pm$2 & GOALS  & SBbc    & N\\
\enddata
\tablecomments{(1): Host galaxy of the nuclear ring. (2)\&(3): coordinates of
the host galaxy. (4): Distance to the host galaxy. For SFRS galaxies, we use
redshift-independent distances reported in \cite{kennicutt11}. For GOALS
galaxies, we use luminosity distances reported in \cite{armus09} where they
adopted the 3-attractor model \citep{mould00}. (5): Total infrared luminosity
(8-1000$\mu$m) of the host galaxy in $\log_{10}$ Solar units
($L_\odot=3.826\times10^{33}\,\mathrm{ergs/s}$), calculated using distances from
(4) and IRAS flux densities in \cite{sanders03} with recipes given in
\cite{sanders96}. NGC 7469 is in a close interacting pair, so we use instead the
value for the resolved major component from \cite{howell10}. (6):
Angular to physical scale conversion factor. For LIRGs, this is
calculated using Ned Wright's Cosmology Calculator \citep{wright06} in
combination with the 3-attractor model \citep{mould00}. (7): References for
active type classification: NGC 1097: \cite{phillips84,s-bergmann93}; NGC 3351:
\cite{planesas97}; NGC 4321:\cite{ho97}; NGC 4826: \citet{garcia03}; NGC 1614:
\cite{herrero17}; NGC 1797:\cite{balzano83}; NGC 7469: \cite{osterbrock93}; NGC
7591: \cite{pereira10}. (8): AGN contribution to the bolometric
luminosity of the host galaxy, estimated by \cite{diaz-santos17} for the GOALS
galaxies using five different mid-IR AGN diagnostics. AGN contribution to the
bolometric light is expected to be negligible in the SFRS galaxies
\citep[e.g.][]{dale09}. (9): Survey with which nuclear ring observations were
taken. (10): Morphology classification of the host galaxy from the NASA/IPAC
Extragalactic Database (NED). (11): Merger stages for the GOALS galaxies are
determined by \cite{stierwalt13} via visual inspection of the Spitzer IRAC
3.6$\mu$m observations (N=non-merger, a=pre-merger, d=late-merger). Using the
same criteria, we classify all SFRS galaxies as non-mergers, given the lack
of massive neighbors in their proximity. $^*$However, NGC 1097 is actively
interacting with at least one dwarf companion \citep{bowen16}.}
\end{deluxetable*}

\begin{deluxetable*}{lllllrcr}[tbh]
\tabletypesize{\scriptsize}
\tablewidth{0pt} 
\tablenum{2}
\tablecaption{Observations and Image Characteristics of VLA Data\label{tab:vlaimaging}}
\tablehead{
\colhead{Galaxy} & 
\colhead{Project ID} &
\colhead{Band} & 
\colhead{Configuration} &
\colhead{Beam} &
\colhead{PA} &
\colhead{Physical Resolution$^a$} &
\colhead{$\sigma_{\mathrm{rms}}$}\\
\colhead{} & 
\colhead{} & 
\colhead{} &
\colhead{} & 
\colhead{} &
\colhead{($^\circ$)} &
\colhead{(pc)} &
\colhead{\tiny{$(\mu\mathrm{Jy/bm})$}}
} 
\startdata 
{ } & 11B-032        & Ka & D  & $3\farcs17\times1\farcs55$ & 26.7  & 107 & 40.4\\
NGC 1097 & 13B-215   & Ku & C  & $3\farcs81\times0\farcs99$ & -7.6  & 69 & 20.0\\
{ } & 13B-215        & S  & B  & $5\farcs76\times1\farcs80$ & -27.4 & 125 & 46.3\\ \hline
{ } & 11B-032        & Ka & D  & $1\farcs76\times1\farcs72$ & -34.3 & 28 & 35.2 \\
IC 342 & 13B-215     & Ku & C  & $1\farcs72\times1\farcs13$ & -9.5  & 18 & 15.1\\
{ } & 13B-215        & S  & B  & $2\farcs23\times1\farcs76$ & -0.8  & 28 & 38.5\\ \hline
{ } &11B-032, 13A-129& Ka & D  & $2\farcs27\times2\farcs04$ & 47.2  & 92 & 17.2 \\
NGC 3351 & 13B-215   & Ku & C  & $1\farcs95\times1\farcs67$ & -17.5 & 75 &14.1\\
{ } & 13B-215        & S  & B  & $2\farcs01\times1\farcs75$ & 43.6  & 79 & 17.2\\ \hline
{ } & 13A-129        & Ka & D  & $2\farcs41\times1\farcs77$ & 42.3  & 123 & 18.5\\
NGC 4321 & 13B-215   & Ku & C  & $1\farcs53\times1\farcs17$ & 17.0  & 82 & 9.7\\
{ } & 13B-215        & S  & B  & $1\farcs89\times1\farcs72$ & 30.6  & 120 & 15.0\\ \hline
{ } & 13A-129        & Ka & D  & $2\farcs16\times1\farcs98$ & 65.9  & 51 & 13.4\\
NGC 4826 & 13B-215   & Ku & C  & $1\farcs46\times1\farcs33$ & 61.3  & 34 &10.4\\
{ } & 13B-215        & S  & B  & $1\farcs97\times1\farcs75$ & 67.3  & 45 & 14.0\\ \hline
{ } &14A-471, 16A-204& Ka & A,C& $0\farcs12\times0\farcs06$ & 16.8 & 19 & 13.5 \\
NGC 1614 &14A-471    & Ku & A,C& $0\farcs17\times0\farcs11$ & -0.4 & 36 & 18.8\\
{ } &14A-471         & S  & A,C& $0\farcs87\times0\farcs55$ & 26.8 & 178 & 12.3\\ \hline
{ } &14A-471         & Ka & C  & $0\farcs76\times0\farcs56^{*}$ & 13.7 & 170 & 16.5\\
NGC 1797 &14A-471    & Ku & A,C& $0\farcs13\times0\farcs12$ & 15.6 & 36 & 10.0\\
{ } &14A-471         & S  & A,C& $0\farcs80\times0\farcs54$ & 7.6 & 164 & 19.4 \\ \hline
{ } &14A-471         & Ka & C  & $0\farcs63\times0\farcs52^{*}$ & 2.4 & 173 & 21.1\\ 
NGC 7469 &14A-471    & Ku & C  & $1\farcs44\times1\farcs12^{*}$ & -30.8 & 372 & 13.5\\
{ } &14A-471         & S  & A,C& $0\farcs65\times0\farcs50$ & -43.4 & 166 & 40.2\\ \hline
{ } &14A-471         & Ka & C  & $0\farcs63\times0\farcs51^{*}$ & -4.1 & 171 & 17.9\\
NGC 7591 &14A-471    & Ku & A,C& $0\farcs13\times0\farcs11$ & -17.6 & 37 & 9.4\\ 
{ } &14A-471         & S  & A,C& $0\farcs68\times0\farcs54$ & -50.7 & 181 & 27.9\\ 
\enddata
\tablecomments{The central frequencies for Ka, Ku and S bands are 33 GHz, 15
GHz, and 3 GHz respectively. The root-mean-square error $\sigma_\mathrm{rms}$
was measured manually on each image before primary beam correction in an
emission-free region. A and C configuration datasets are weighted with
100:1 (A:C) before combined imaging to account for differences in $uv$-plane
distribution.} 
\tablenotetext{*}{C configuration image was used due to limited
dynamic range at A configuration.} \tablenotetext{a}{The smallest physical scale
resolved by the synthesized beam at the distance of the galaxy, given
the FWHM of the minor axis of the beam.}
\end{deluxetable*}
    
\begin{deluxetable*}{ccccccc}
\tabletypesize{\scriptsize}
\tablewidth{0pt} 
\tablenum{3}
\tablecaption{Observations and Image Characteristics of Ancillary CO (J=1-0) Data}\label{tab:almaimaging}
\tablehead{
\colhead{Galaxy} & 
\colhead{Telescope} &
\colhead{Project ID/Reference} & 
\colhead{Beam} &
\colhead{PA} &
\colhead{Physical Resolution $^a$} &
\colhead{$\sigma_\mathrm{rms}$}\\
\colhead{} & 
\colhead{} & 
\colhead{} & 
\colhead{} &
\colhead{($^\circ$)} &
\colhead{(pc)} &
\colhead{\tiny{$(\mathrm{mJy/bm\ km/s})$}}
} 
\startdata 
NGC 1097 & ALMA &2012.1.00001.S & $2\farcs85\times2\farcs57$ & 72.6 & 178 & 2.8\\
IC  342  & NRO10m &  \cite{ishizuki90}     &  $2\farcs39\times2\farcs27$  &-31  & 36  & 3.5\\
NGC 3351 & ALMA & 2013.1.00885.S & $2\farcs03\times1\farcs22$ & 27.8 & 55 & 0.5\\
NGC 4321 & ALMA & 2015.1.00978.S  & $2\farcs40\times2\farcs24$ & -31.0 & 156 & 0.4\\ 
         & ALMA & 2016.1.00972.S & & &\\
NGC 4826 & IRAM & \cite{casasola15}  &  $2\farcs53\times1.\farcs80$ &  39.0 & 156 &  0.25\\ 
NGC 7469 & ALMA & 2013.1.00218.S & $0\farcs91\times0\farcs51$ & -48.8 & 169 & 0.2\\
\enddata
\tablecomments{For NGC 4321, we combined the measurement sets from two different
ALMA projects to produce CO(J=1-0) moment 0 maps. For ALMA datasets, the
root-mean-square error $\sigma_\mathrm{rms}$ was measured manually on each
continuum-subtracted line cube before primary beam correction in an
emission-free region across all channels. For NGC 4826 and IC 342,
$\sigma_\mathrm{rms}$ were taken from the original references. ALMA
datasets for NGC 1097 and NGC 4321 are unpublished. Dataset for NGC 3351 is
published in \cite{leaman19}, and dataset for NGC 7469 is published in
\cite{zara17} and \cite{wilson19}.}
\tablenotetext{a}{The smallest physical scale resolved by the synthesized beam at the distance of the galaxy, given the FWHM of the minor axis of the beam.}
\end{deluxetable*}

\section{Sample Selection} \label{sec:sample} The GOALS ``equatorial" VLA
campaign \citep[][; Song et al. in prep]{linden19} is a multi-frequency,
multi-resolution snapshot survey designed to map the brightest radio continuum
emission in all 68 U/LIRGs from GOALS that are within declination $|\delta| <
20^\circ$ at Ka (33 GHz), Ku (15 GHz) and S (3 GHz) bands at kpc and
$\sim$ 100\,pc scales. These observations serve as a companion to SFRS, which
imaged 56 nearby ($D_L < 30$\,Mpc) normal star-forming galaxies (i.e.
non-U/LIRGs) from SINGS \citep{kennicutt03} and KINGFISH \citep{kennicutt11} at
matched frequencies and also at both kpc \citep{murphy12} and $\sim$
100\,pc scales \citep{murphy18,linden20}. Using the kpc resolution observations
from the GOALS equatorial survey, \cite{linden19} studied extra-nuclear star
formation in 25 local LIRGs, and concluded that the high global SFR of
these systems, relative to the Star Formation Main Sequence
\citep[e.g.][]{elbaz11,speagle14} occupied by normal galaxies in SFRS, must be
driven by extreme nuclear SF.\\
\indent In this work we focus on studying and comparing star-forming
properties of nuclear rings at $\sim$ 100\,pc scales in a sample of nine
galaxies from SFRS and GOALS. While the term ``nuclear ring'' is traditionally
reserved for rings forming at the Nuclear Lindblad Resonance \citep{fukuda98}, it
has also been more broadly used to describe the innermost star-forming rings in
galaxies \citep[e.g.][]{boeker08}. In this paper, we follow the definition and
size measurements given in a study of 113 nearby nuclear rings by
\cite{comeron10}, and use ``nuclear ring" to describe ring-like emission within
the central 2\,kpc of a galaxy detected in our surveys at 3, 15 or 33 GHz in
radio continuum. With the exceptions of the rings in NGC 1797 and NGC 7591,
which are resolved for the first time with the GOALS equatorial survey, all other
rings reported here have previously been identified and separately studied as
``nuclear rings" at various wavelengths \cite[e.g.][]{mazzuca08, comeron10, ma18}.
In some other studies, these rings have also been referred to as ``circumnuclear
rings'' \citep[e.g.][]{Xu15,prieto19}. Figure \ref{fig:natres1} and
\ref{fig:natres2} show the VLA images of the sample, and Table \ref{tab:sample}
lists the basic properties of the host galaxies.\\
\indent We note that due to the sensitivity limit of our VLA observations, we
were only able to identify the nine nuclear rings that show consistently bright
($\gtrsim 0.1$mJy/beam) high-frequency (15 or 33 GHz) emission among the
combined total of 124 targets from SFRS and the GOALS equatorial survey. Several
other galaxies, such as NGC 1068 (LIRG) and NGC 5194 (normal), are also known
nuclear ring hosts from optical studies \citep[e.g.][]{Telesco88,lauri02}, but
their ring structures were not detected in our observations. We further discuss
the potential biases of the sample selection in Section \ref{dis:bias}. 


\section{Observations \& Data Reduction}\label{sec:vladata} \subsection{VLA
data} Each galaxy was observed with the VLA at S (3 GHz), Ku (15 GHz) and Ka
Band (33 GHz). Observations of the five normal galaxies are acquired from the
SFRS. This survey observed all galaxies at S, Ku, and Ka Band in B, C and
D configuration, respectively, to achieve a common angular resolution of $\sim
2\farcs0$ at each frequency. Details of SFRS observations and data calibration procedures are
described in \cite{murphy18} and \cite{linden20}.\\
\indent The four LIRGs were observed with both A ($ 0\farcs06 - 0\farcs6$) and C
($0\farcs6 - 7\farcs0$) configurations at 3, 15 and 33 GHz. The observations
were carried out with projects VLA/14A-471 (PI: A. Evans) and VLA/16A-204 (PI:
S. Linden). The raw VLA datasets were first reduced and calibrated with the
standard calibration pipelines using the Common Astronomy Software Applications
\citep[CASA, v4.7.0;][]{mcmullin07}. Each measurement set was visually
inspected and data related to bad antennas, time and frequency ranges (including
RFI) were manually flagged. The above two steps were repeated until all bad data
were removed.\\
\indent Before imaging, we performed data combination on the reduced A and C
configuration measurement sets at each frequency using CASAv4.7.0 \texttt{concat}
task with a weighting scale of 100:1 (A:C). This was done to enhance the image
sensitivity while maximizing resolution, as well as maintaining good PSF shapes
(i.e. Gaussian-like), accounting for the fact that the \textit{uv}-plane
distribution of C configuration data is $\sim$ 100 times denser than A
configuration. The concatenated measurement sets
were then imaged using CASAv4.7.0 \texttt{tclean} task, using Briggs weighting
with a robust parameter of 1.0 and (Multi-Scale) Multi-Frequency Synthesis
deconvolving algorithm. \\
\indent Due to short on-source times, high resolution A configuration
imaging is unable to recover the nuclear ring emission at 33 GHz for NGC 1797
and NGC 7591, and at both 15 and 33 GHz For NGC 7469. Therefore, in these cases,
we use images made from C configuration measurement sets only, using the
same \texttt{tclean} parameters as above. Additionally, bright nuclear emission
in NGC 7469 allowed for self-calibration of the C Configuration data at 3, 15
and 33 GHz. Native resolution images of all data referred in this work are
displayed in Figure \ref{fig:natres1} and \ref{fig:natres2}, and descriptions of
these images and extra information on individual sources are provided in
Appendix \ref{sec:sample_detail}. The characteristics of these images and
associated observation information are listed in Table
\ref{tab:vlaimaging}. 

\subsection{Archival CO(J=1-0) Data}\label{sec:almadata} To better understand
the properties of star formation in nuclear rings, it is informative to study
the molecular environments that give rise to this activity. For each host galaxy
in the sample, we searched for archival CO(J=1-0) observations with resolutions
that are comparable to or higher than our VLA observations to directly compare
SF and cold molecular gas properties in its nuclear ring. Four galaxies in our
sample have publicly available archival ALMA data that meet the above criteria,
from ALMA projects 2012.1.00001.S (NGC 1097, PI: K. Sheth; unpublished),
2013.1.00885.S \citep[NGC 3351, PI: K. Sandstrom;][]{leaman19}, 2015.1.00978.S
(NGC 4321, PI: K. Sandstrom; unpublished), 2016.1.00972.S (NGC 4321, PI: K.
Sandstrom; unpublished) and 2013.1.00218.S \citep[NGC 7469, PI: T.
Izumi;][]{zara17, wilson19}. Each ALMA dataset was first re-processed using the
appropriate CASA version and data calibration pipeline that are specified in the
project's data calibration note from the archive. Continuum subtraction was
performed using the \texttt{uvcontsub} task in CASA (version consistent with
reduction pipeline). The reduced measurement sets were then re-imaged
using \texttt{tclean} in CASAv4.7.0, using Briggs weighting with robust
parameter of 0.5 and H\"{o}gbom cleaning algorithm. For NGC 1097, because the
native angular resolution of the dataset is about four times higher than the VLA
data, we also tapered the \textit{uv}-distribution of the measurement
sets with a 3$\farcs$0 Gaussian kernel to better compare with the VLA data. We
then produced moment 0 maps from the continuum-subtracted CO(J=1-0) line cubes
using the \texttt{immoments} task in CASAv4.7.0.\\
\indent Additionally, we downloaded high-resolution CO (J=1-0) moment 0 maps for
NGC 4826 \citep{casasola15} and IC 342 \citep{ishizuki90} from NED. The map for
NGC 4826 was observed by the IRAM Plateau de Bure Interferometer (PdBI), and
provided in units of M$_\odot$\,pc$^{-2}$. We used information from Table 1 and
Equation 2 in \cite{casasola15} to convert the data to units of Jy/beam km/s.
The map for IC 342 was observed using the 10-m sub-millimeter telescope (NRO10m)
at Nobeyama Radio Observatory, and provided in units of Jy/beam m/s in B1950
coordinates. We used the \texttt{imregrid} task in CASA to re-map the datasets
to match the coordinates of our VLA data (J2000), after which we converted the
data to units of Jy/beam km/s. \\
\indent Characteristics of these moment 0 maps are listed in
Table \ref{tab:almaimaging}. In Section \ref{sec:tdep_analysis}, we utilize these moment
0 maps to estimate the cold molecular gas densities and gas depletion times in
these six nuclear rings.

\renewcommand{\thefigure}{\arabic{figure}}
\addtocounter{figure}{+1}
\begin{figure*}[thb!]
    \centering
    \epsscale{1.2}
    \plotone{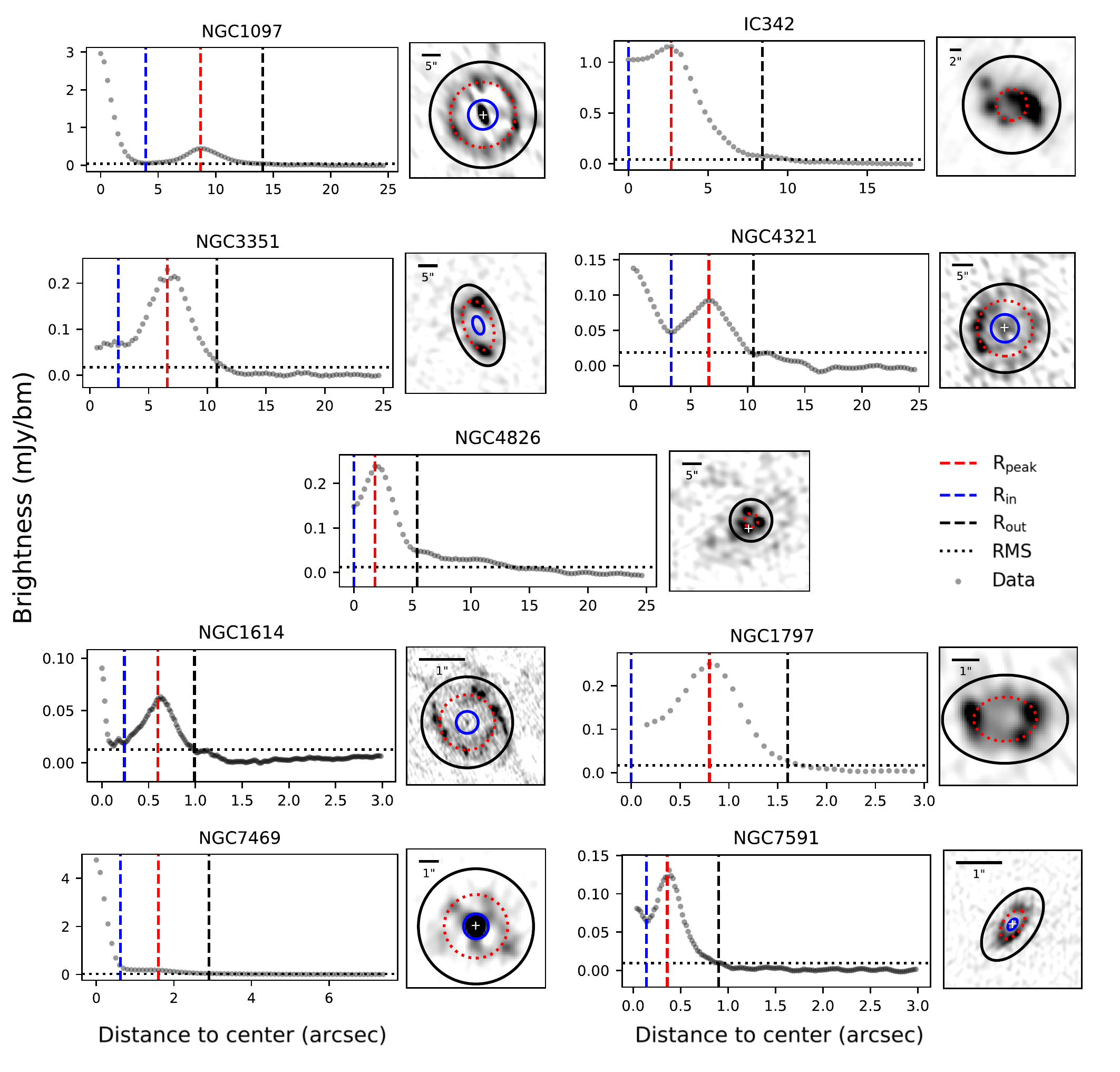} 
    \caption{\textit{Left column}: azimuthally-averaged
    light profile (in grey dotted line) from the 33 GHz image of each nuclear ring in
    the sample. The red dashed vertical line shows the radius of the ring
    ($R_\mathrm{peak}$), and the blue and black dashed vertical lines mark the
    inner and outer radius ($R_\mathrm{in}$, $R_\mathrm{out}$), respectively.
    See Section \ref{sec:integrated} for definitions of the different radii.
    \textit{right column}: 33 GHz image of the nuclear ring shown in grey scale.
    Circles/ellipses overlaid in dotted red, solid blue and black mark
    $R_\mathrm{peak}$, $R_\mathrm{in}$ and $R_\mathrm{out}$, respectively. White ``+" mark the
    locations of AGN. For NGC 7591, observation and measurements at 15 GHz are
    used here because the available 33 GHz image does not resolve the ring
    structure. \label{fig:lightprofile}}
\end{figure*}
\begin{figure*}[tbh!]
    \epsscale{0.9}
    \plotone{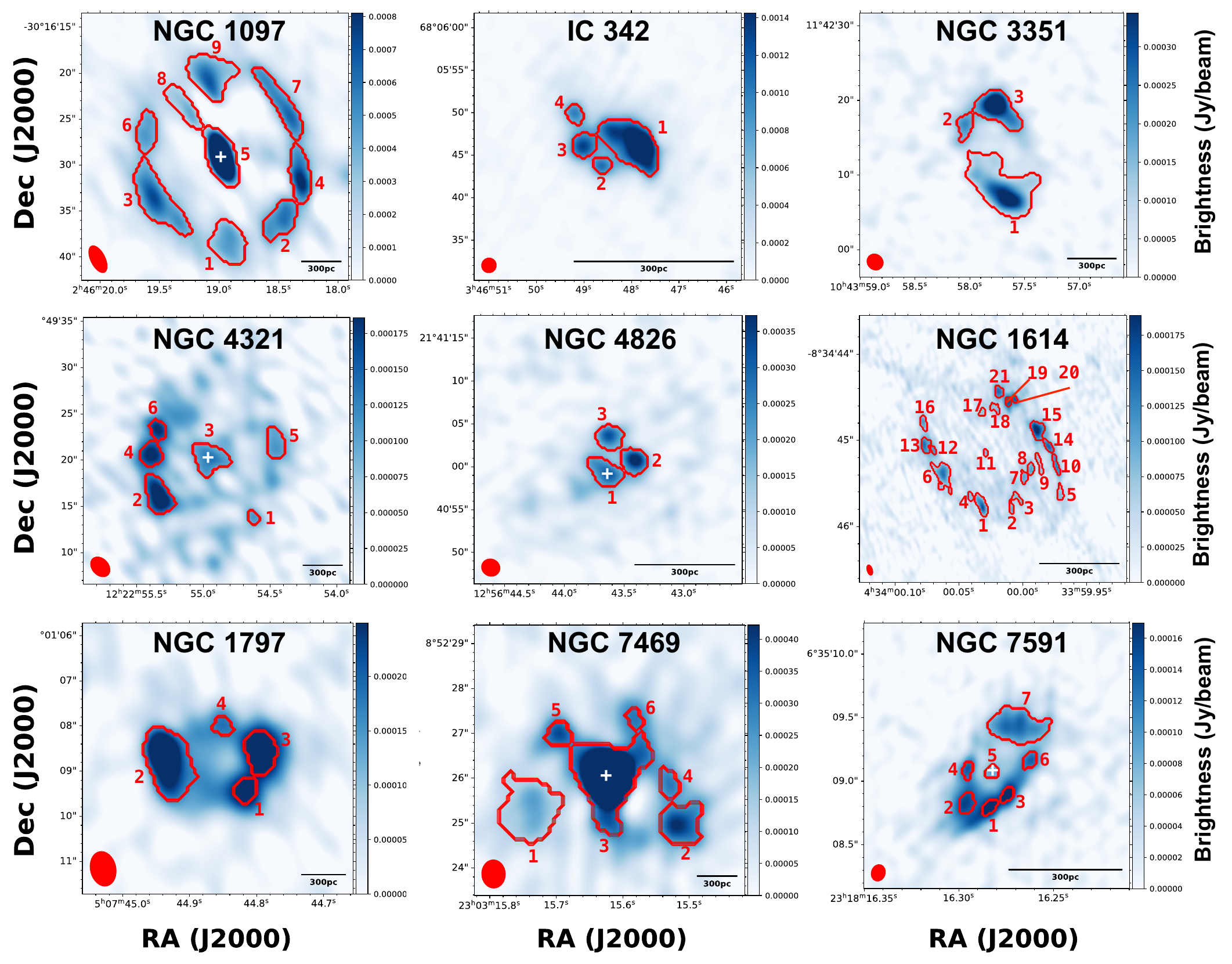}
    \caption{33 GHz images of the sample galaxies (15 GHz for NGC 7591) with
    nuclei and NRSF regions identified with \textit{Astrodendro} outlined in red. Region IDs
    (labeled in red) are assigned in ascending order based on region
    declinations. Red ellipses on the bottom left of the images represent the
    shapes and sizes of synthesized beams. Scale bars shown on the lower right
    reflect the physical scales of the observed structures at the distance of
    the host galaxy. White ``+'' signs mark the locations of known AGN.
    \label{fig:dendro_sfr}}
\end{figure*}

\begin{deluxetable*}{cccccllll}[tbh!]
\tablecaption{Integrated Nuclear Ring Properties\label{tab:integrated}}
\tablecolumns{9}
\tablenum{4}
\tablewidth{0pt}
\tablehead{
\colhead{Galaxy} &
\colhead{$\nu$(GHz)} & 
\colhead{$R_\mathrm{in}$(pc)} &
\colhead{$R_\mathrm{peak}$(pc)} &
\colhead{$R_\mathrm{out}$(pc)} &
\colhead{$S^r_\nu$(mJy)} &
\colhead{SFR(M$_\odot$ yr$^{-1}$)} &
\colhead{$\Sigma_\mathrm{SFR}$(M$_\odot$ yr$^{-1}$ kpc$^{-2}$)} &
\colhead{SFR$_\mathrm{ring}$/SFR$_\mathrm{tot}$(\%)}\\
\colhead{(1)} &
\colhead{(2)} &
\colhead{(3)} &
\colhead{(4)} &
\colhead{(5)} &
\colhead{(6)} &
\colhead{(7)} &
\colhead{(8)} &
\colhead{(9)}
}
\startdata  
NGC 1097 & 33 & 269 & 599 & 970  & 19$\:\pm$2.0    & 2.0$\;\,\pm$0.20  & 0.73$\pm$0.07  & 39$\pm$4\\
IC 342   & 33 & -   & 43   & 134   & 29$\:\pm$2.9    & 0.16$\pm$0.02 & 2.90$\pm$0.29  & 7$\;\,\pm$1 \\ 
NGC 3351 & 33 & 109 & 299  & 489   & 4.5$\pm$0.46    & 0.20$\pm$0.02  & 0.51$\pm$0.05 & 24$\pm$2\\
NGC 4321 & 33 & 229 & 458  & 728  & 3.9$\pm$0.42    & 0.41$\pm$0.04  & 0.27$\pm$0.03  & 13$\pm$1\\
NGC 4826 & 33 & -   & 46   & 138   & 2.5$\pm$0.26    & 0.03$\pm$0.01  & 0.69$\pm$0.07  & 7$\;\,\pm$1\\      
NGC 1614 & 33 & 77  & 194  & 319   & 12$\:\pm$1.3    & 29$\;\;\,\pm$3.0     & 97\;\;\,$\pm$9.9    & 60$\pm$8\\
NGC 1797 & 33 & -   & 242  & 485   & 3.0$\pm$0.30    & 6.1$\;\,\pm$0.63  & 6.0\;\,$\pm$0.62  & 49$\pm$5\\
NGC 7469 & 33 & 208 & 531  & 963   & 8.1$\pm$0.83    & 21$\;\;\,\pm$2.1    & 7.6\;\,$\pm$0.78  & 57$\pm$7\\
NGC 7591 & 15 & 47  & 121  & 301   & 4.2$\pm$0.43    & 8.0$\;\,\pm$0.82  & 53\;\;\,$\pm$5.3    & 59$\pm$6\\
\enddata
\tablecomments{(1): Host galaxy of the nuclear ring. (2): Frequency at which
sizes and SFR are measured. (3)-(5): inner, peak and outer radius of the ring,
as defined in Section \ref{sec:integrated}. Details on individual source are
described in Appendix \ref{sec:sample_detail}.(6): Flux density enclosed by
apertures with $R_\mathrm{out}$ and $R_\mathrm{in}$. For NGC 4826, off-centered
AGN emission is excluded. (7): Star formation rates calculated using
Eq.\ref{eq_sfr} and values from (6). (8): Star formation rate surface densities
of each ring, calculated by dividing (7) over the area the ring extends. See
Section \ref{sec:ring} for details. (9): Fraction of the total SFR of the
host galaxy in the nuclear ring. See Section \ref{sec:ring_frac} and Table \ref{tab:host_sfr} for details. 
}
\end{deluxetable*}

\begin{deluxetable}{llll}[tbh!]
\tablecaption{Host Galaxy Star Formation Rates}
\tablecolumns{4}
\tablenum{5}
\tablewidth{0pt}
\tablehead{
\colhead{Galaxy} &
\colhead{SFR$_\mathrm{FUV}$} & 
\colhead{SFR$_\mathrm{IR}$} &
\colhead{SFR$_\mathrm{tot}$} \\
\colhead{} &
\colhead{(M$_\odot$ yr$^{-1}$)} &
\colhead{(M$_\odot$ yr$^{-1}$)} &
\colhead{(M$_\odot$ yr$^{-1}$)}  \\
\colhead{(1)} &
\colhead{(2)} &
\colhead{(3)} &
\colhead{(4)} 
}
\startdata  
NGC 1097 & 0.69$\pm$0.03 & 4.4$\pm$0.0  & \;\:5.1$\pm$0.0\\
IC 342   & 1.50$\pm$0.07  & 0.9$\pm$0.0 & \;\:2.4$\pm$0.0 \\ 
NGC 3351 & 0.15$\pm$0.01 & 0.7$\pm$0.0 & 0.86$\pm$0.01\\
NGC 4321 & \;\:0.7$\pm$0.03  & 2.6$\pm$0.1  & \;\:3.3$\pm$0.1\\
NGC 4826 & 0.04$\pm$0.01 & 0.4$\pm$0.0 & 0.46$\pm$0.01\\      
NGC 1614 & \;\:1.5$^{*}$ & \;47$\pm$4.3   & \;\;\:48$\pm$4\\
NGC 1797 & \;\;\;\;\;-$^{*}$ & \;12$\pm$0.4   & \;\;\:12$\pm$0.4\\
NGC 7469 & \;\:1.7$\pm$0.0   & \;37$\pm$3.0   & \;\;\:39$\pm$3 \\
NGC 7591 & \;\:0.4$\pm$0.1   & \;13$\pm$0.3   & \;\;\:14$\pm$0.3 \\
\enddata
\tablecomments{(1): Host galaxy name. (2): Star formation
rates derived from GALEX FUV flux measurements by \cite{clark18},
\cite{howell10} and \cite{brown14}, using Eq.(3).$^*$GALEX measurements are not
available for NGC 1614 and NGC 1797. For NGC 1614, we use monochromatic UV SFR
from \cite{u12} scaled down by a factor of two as an estimate for FUV SFR. (3):
Star formation rates derived from total IR luminosity, accounting for
AGN contribution to the bolometric luminosity of the LIRGs (Table
\ref{tab:sample}), using Eq.5. (4): Total star formation rates of the host
galaxy. See Section \ref{sec:ring_frac} for more details.\label{tab:host_sfr}}
\end{deluxetable}
\section{Data Analysis}\label{sec:analysis} Here we mainly use the 33 GHz
continuum images for our analysis of the nuclear rings and individual NRSF
regions because radio continuum emission above 30 GHz has been shown to
effectively trace thermal free-free emission associated with ongoing massive SF
in both normal galaxies and LIRGs \citep[e.g.][]{murphy12,linden19,linden20}.
For NGC 7591, we use the 15 GHz image instead because the available 33 GHz image
does not resolve the ring structure. 
\subsection{Integrated ring measurements}\label{sec:integrated}
\indent To measure the integrated ring properties, we first quantify the spatial
extent of each ring by defining an inner, peak, and outer radius/semi-major axis
($R_\mathrm{in}$, $R_\mathrm{peak}$, $R_\mathrm{out}$) based on
its azimuthally-averaged light profile, as shown in Figure \ref{fig:lightprofile}.
These light profiles are measured from the central coordinates of the host
galaxies using 1-pixel wide circular annuli. Elliptical annuli are used for highly
elliptical rings (NGC 3351, NGC 1797, NGC 7591).  Details on the relevant
procedures are provided in Appendix \ref{sec:integrated_method}. In general, we
locate $R_\mathrm{in}$ at the first local minimum of the light profile, and
define $R_\mathrm{peak}$ at the local maximum outside of $R_\mathrm{in}$. To
account for diffuse emission from the ring that is not necessarily
axis-symmetric, we define $R_\mathrm{out}$ to be where the averaged light
profile flattens towards the image noise level (i.e. $\sigma_\mathrm{rms}$ in
Table \ref{tab:vlaimaging}). An exception to this is NGC 4826, whose averaged
light profile contains contribution from a faint spiral structure that surrounds
the ring, which we exclude from further analysis by setting $R_\mathrm{out}$ at
the local minimum immediately outside $R_\mathrm{peak}$. Due to the limited
resolution of the observations, light profiles of IC 342, NGC 4826 and NGC 1797
do not yield a well defined local minimum, therefore we do not use
$R_\mathrm{in}$ for rings in these three galaxies. See Appendix
\ref{sec:sample_detail} for more details on individual sources.\\
\indent The flux of each ring is then measured within the area characterized by
$R_\mathrm{in}$ and $R_\mathrm{out}$ via aperture photometry. For NGC 4826, a
LINER AGN likely contributes to the ring emission due to $m=1$ perturbation
\citep{garcia03}, therefore we additionally mask the image at the reported AGN
location ($\alpha_\mathrm{J2000}=$12h56m43.64s, $\delta_\mathrm{J2000}=$21d40m
59.30s) with a beam-sized aperture before performing aperture photometry. In
Figure \ref{fig:lightprofile}, we also mark the locations of known AGN with
``+". We do not find similar cases of AGN emission contributing to the nuclear
ring in the rest of the sample.
\subsection{NRSF Region identification \& measurements}\label{sec:sfregions} Given
the high resolution of our VLA observations, all nine nuclear rings are resolved at
sub-kpc scales at 33 GHz or 15 GHz, thus allowing us to further characterize the
properties of individual star-forming regions in these rings. \\
\indent To identify and measure the flux of individual NRSF regions in each
nuclear ring, we use the software \textit{Astrodendro} \citep{astrodendro},
which measures hierarchical structures in an astronomical image using
dendrograms. \textit{Astrodendro} identifies and categorizes structures in an
image into \texttt{trunk}, \texttt{branch} and \texttt{leaf}, based on three
input parameters: the minimum brightness required for a structure to be
physically meaningful (\texttt{min\_value}), the minimum number of pixels in a
structure (\texttt{min\_npix}), and the minimum brightness relative to the local
background required for a structure to be considered independent
(\texttt{min\_delta}). Structures identified as \texttt{leaf} are of the highest
hierarchical order, which are the individual NRSF regions that we are interested
in, while \texttt{branch} and \texttt{trunk} are the less luminous diffuse ring
emission connecting the SF regions. Therefore here we only focus on the derived
properties of \texttt{leaf} structures.\\
\indent We run \textit{Astrodendro} on the 33 GHz image of each nuclear ring (15
GHz for NGC 7591) with \texttt{min\_value}$=5\sigma_\mathrm{rms}$ and
\texttt{min\_delta}=$1\sigma_\mathrm{rms}$ to identify physically meaningful
\texttt{leaf} structures, where $\sigma_\mathrm{rms}$ is the rms noise measured
in an emission-free region of the image before primary beam correction (see
Table \ref{tab:vlaimaging}). We set \texttt{min\_npix} to be 1/4 the area of the
synthesized beam, to avoid identifying noise spikes yet allowing detection of
small, unresolved regions. Afterwards, we manually eliminate regions identified
beyond the outer radius ($R_\mathrm{out}$) of the ring, to ensure we only
include NRSF regions or nuclei in subsequent analyses. Figure
\ref{fig:dendro_sfr} shows all the identified regions using
\textit{Astrodendro}. We note that varying \texttt{min\_npix} by small
amounts does not significantly alter the population of identified \texttt{leaf}
structures. For example, setting \texttt{min\_npix} to be  1/2 or 3/4 of the
beam area only blends together Region 1 and 3 in NGC 1797, and Region 3 and 6 in
NGC 7469, and the rest of region  
identification remains exactly the same, in terms of numbers, sizes and shapes.\\
\indent The flux density and angular area of each identified region are also
measured by \textit{Astrodendro}, which can be heavily dependent on the
signal-to-noise level of the region. To characterize the effect of image noise
on the region size and flux measurements, we use a Monte Carlo method and
re-run \textit{Astrodendro} 1000 times randomly adjusting the brightness of each pixel
sampling from a Gaussian distribution defined by the rms noise
$\sigma_\mathrm{rms}$ and the assumed VLA flux calibration error ($\sim$ 10\%).
The standard deviations of the results from the 1000 runs are used to quantify
the uncertainties in the flux and size measurements. For unresolved
regions that have \textit{Astrodendro}-measured sizes smaller than the beam areas after accounting
for uncertainties, we instead measure their flux using beam-sized apertures and
use the beam areas as upper-limits for their sizes.\\
\indent Additionally, to estimate the ratio of thermal free-free emission to
total radio continuum emission in these NRSF regions at 33 GHz, we measure radio
spectral index between 15 and 33 GHz associated with each region. To do so, we
smooth and regrid the native resolution 15 GHz and 33 GHz images of each nuclear
ring (shown in Figure \ref{fig:natres1} and \ref{fig:natres2}) to a common
circular beam and pixel scale for consistent measurements of flux densities
across the two frequencies. Assuming a single power-law model representing the
combination of flat-spectrum thermal emission and steep-spectrum non-thermal
emission $S \sim \nu^{\alpha}$, we can calculate the spectral index $\alpha$
associated with each region by measuring the linear slope between flux densities
measured at 15 and 33 GHz with respect to frequency. Due to the coarser
resolutions of the beam-matched images, the exact region areas identified with
\textit{Astrodendro} using native resolution images cannot be used here to
extract spectral indices. As an approximation, for each region we measure its
beam-matched 15 and 33 GHz flux using a common circular aperture with area
equivalent to the region size as measured by \textit{Astrodendro}. For regions
with sizes smaller than the matched circular beams, we instead use beam-sized
apertures to extract their associated spectral indices. Uncertainties in the spectral indices
are calculated with error propagation. 
\subsection{Measurements of CO(J=1-0) maps}\label{sec:tdep_analysis} For the six galaxies
that have high-resolution ancillary CO(J=1-0) data (see Section
\ref{sec:almadata}), we smooth and regrid the native resolution CO (J=1-0) moment
0 map and 33 GHz radio continuum map of the nuclear ring to a common circular
beam and pixel scale for consistent flux measurements. We measure the total CO
(J=1-0) and 33 GHz continuum flux of each ring using apertures defined by
$R_\mathrm{in}$ and $R_\mathrm{out}$ (See Section \ref{sec:integrated}). These
values are used to derive the integrated molecular gas mass, average surface
densities and gas depletion times in the ring in Section \ref{sec:tdep_result}.
For individual NRSF regions, we measure the CO (J=1-0) and 33 GHz flux using
circular apertures with area equivalent to the region size as measured by
\textit{Astrodendro}. For regions with sizes smaller than the matched circular
beams, we use beam-sized apertures instead.

\section{Results}\label{sec:results} Here we describe the results of the above
measurements and further derive SFR, SFR surface density and gas depletion time for each ring as
a whole, as well as for individual NRSF regions identified using
\textit{Astrodendro}. We also use the measured 15 -- 33 GHz spectral index to
derive the ratio of thermal free-free emission to total radio emission at 33 GHz
associated with each NRSF regions. The measured and derived quantities for the
nuclear rings are summarized in Table \ref{tab:integrated}, and the ones for
individual NRSF regions are reported in Table \ref{tab:resolved}. In total, 63
regions are identified and measured with \textit{Astrodendro}, as shown in
Figure \ref{fig:dendro_sfr}, including 58 NRSF regions (22 in normal
galaxies, and 35 in LIRGs) and 5 nuclei (4 known AGN and 1 nuclear starburst).
For completeness, in Table \ref{tab:resolved} we report measurements for all 63
identified regions, but only the 58 NRSF regions are included in the final
analysis. 
\subsection{Ring size, SFR \& SFR surface density}\label{sec:ring}
\indent We use $R_\mathrm{peak}$, defined in Section \ref{sec:integrated}, as an
estimate for the radius/semi-major axis of the ring.  The five nuclear rings in
the sample of normal galaxies have radii of 43 -- 599\,pc, and the four
nuclear rings in the LIRGs have radii of 121 -- 531\,pc. \\
\indent Using flux measured with $R_\mathrm{in}$ and $R_\mathrm{out}$, we can
calculate the integrated SFR within each ring using Equation 10 in
\cite{murphy12}:
\begin{align}\label{eq_sfr}
     \Big(\frac{\mathrm{SFR}}{\textup{M}_\odot \mathrm{yr}^{-1}}\Big)  & = 10^{-27}\Big[2.18\Big(\frac{T_e}{10^4 \mathrm{K}}\Big)^{0.45}\Big(\frac{\nu}{\textup{GHz}}\Big)^{-0.1} + \notag\\ 
    & 15.1\Big(\frac{\nu}{\textup{GHz}}\Big)^{\alpha^{\textup{NT}}}\Big]^{-1}\Big(\frac{L_\nu}{\textup{erg} \textup{s}^{-1} \textup{Hz}^{-1}}\Big)
\end{align} 
where a Kroupa Initial Mass Function (IMF) is assumed. In this equation, $L_\nu$
is the spectral luminosity at the observed frequency $\nu$, given by $L_\nu =
4\pi D_L^2 S^r_\nu$, where $S^r_\nu$ is the measured total flux of the ring in
Jy, $D_L$ is the luminosity distance of the host galaxy (column 4 in Table
\ref{tab:sample}), $T_e$ is the electron temperature and $\alpha^{\mathrm{NT}}$
is the non-thermal spectral index. Here we adopt $T_e=10^4$\,K and
$\alpha^{\mathrm{NT}}=-0.85$, which have been extensively used to describe SF
regions in normal galaxies and LIRGs
\citep[e.g.][]{murphy12,linden19,linden20}.\\
\indent The integrated SFR has a range of 0.03 -- 2.0 and 6.1 -- 29 \,M$_\odot$
yr$^{-1}$ for rings in the normal galaxies and in the LIRGs, respectively. For
the normal galaxies, the estimated nuclear ring radii and SFR are in agreement
with previous measurements of the same galaxies at optical and IR wavelengths
\citep[e.g.][]{mazzuca08,comeron10,hsieh11,ma18}. Only a handful of similar
measurements exist for the nuclear rings hosted in LIRGs because they are
farther away and more obscured by dust. Our radio measurement is also consistent
with extinction-corrected Pa$\alpha$ measurement for the nuclear ring in NGC
1614 by \cite{alonso01}, which confirms the effectiveness of using
high-frequency radio continuum as extinction-free SFR tracer in these nuclear
rings.\\
\indent We further calculate the average SFR surface density,
$\Sigma_\mathrm{SFR}$, in each ring, by dividing the integrated SFR over the area
of the ring as defined in Figure \ref{fig:lightprofile}, with $R_\mathrm{in}$
and $R_\mathrm{out}$. For rings with undefined $R_\mathrm{in}$ due to lack of
resolution, we use the areas defined by their $R_\mathrm{out}$ minus the areas
of the synthesized beams (Table \ref{tab:vlaimaging}) to account for the central
cavities. The resulted range of $\Sigma_\mathrm{SFR}$ is 0.27 -- 2.90\,M$_\odot$
yr$^{-1}$ kpc$^{-2}$ for nuclear rings in the normal galaxies, with a median
value of 0.59$\pm$0.21\,M$_\odot$ yr$^{-1}$ kpc$^{-2}$. For rings in the LIRGs,
$\Sigma_\mathrm{SFR}$ is higher by at least a factor of two, with a range of 6.0 --
97\,M$_\odot$ yr$^{-1}$ kpc$^{-2}$ and a median of 30$\pm$22\,M$_\odot$
yr$^{-1}$ kpc$^{-2}$. 
\subsection{Ring SFR vs. Host SFR}\label{sec:ring_frac}
\indent Here we estimate the fraction of total SFR of the host galaxy
contributed by the nuclear ring. The relevant results are tabulated in
Table \ref{tab:integrated} and \ref{tab:host_sfr}. The total SFR of the galaxy is
calculated from both FUV and IR emission to account for obscured and unobscured
SF \citep{murphy12}:
\begin{equation}
    \mathrm{SFR_{tot}}=\mathrm{SFR_{FUV}}+\mathrm{SFR_{IR}}
\end{equation}
To calculate $\mathrm{SFR_{FUV}}$, we use GALEX FUV measurements from
\cite{clark18} for the normal galaxies and from \cite{howell10} and
\cite{brown14} for the LIRGs, along with Equation (2) from
\cite{murphy12} assuming Kroupa IMF:
\begin{equation}
    \mathrm{SFR_{FUV}}=4.42\times10^{-44}L_\mathrm{FUV}
\end{equation}
No GALEX FUV measurements are available for NGC 1614 and NGC 1797. For
NGC 1614, SFR based on monochromatic UV measurement ($\lambda = 2800$\r{A}) is available from
\cite{u12}. Due to the different calibrations adopted, UV SFR from \cite{u12} are
consistently higher than values estimated using FUV measurements from
\cite{howell10} among U/LIRGs studied in both works, by at least a factor of
two. Therefore, for NGC 1614, here we adopt the SFR reported in \cite{u12},
but scaled down by a factor of two as an estimate for its FUV SFR. The FUV
contribution to the total SFR is overall very low ($\sim$ 4\%) in local U/LIRGs
\citep{howell10} therefore does not affect our estimates significantly. For the
IR component, we applied $L_\mathrm{IR}$ from Table \ref{tab:sample} to Equation
(15) in \cite{murphy12}, modified to account for AGN emission:
\begin{align}
    \mathrm{SFR_{IR}}&=3.15\times10^{-44}L_\mathrm{IR,SF}\\
                     &=3.15\times10^{-44}L_\mathrm{IR}(1-f_\mathrm{AGN})
\end{align}
where $f_\mathrm{AGN}$ is the fraction of the bolometric luminosity of the host
galaxy contributed by AGN emission \citep[see Table
\ref{tab:sample} and][]{diaz-santos17}. Because the above relations presented
in \cite{murphy12} are calibrated against 33 GHz measurements, we can directly
estimate the fraction of total SFR contributed by the nuclear ring by dividing
the ring SFR, derived in the last Section, over SFR$_\mathrm{tot}$. The
fractions are 7\% -- 39\% for rings in the normal galaxies, and 49 -- 60\% for
rings in the LIRGs, with median values of 12$\pm$9\% and 56\%$\pm$6\%,
respectively. We visualize this result in Figure \ref{fig:ringfraction}
and discuss its implication in Section \ref{dis:ringsfr}. For NGC 1614, a
similar fraction has also been estimated by \cite{Xu15}. Even though
measurements are made at 15 GHz for NGC 7591, we do not expect new measurements
at 33 GHz to significantly alter our result. 
\subsection{Region size, SFR and
SFR surface density}\label{sec:astrodendro} For each identified NRSF region, we
calculate its SFR and $\Sigma_{\mathrm{SFR}}$ using the flux density and area
measured in Section \ref{sec:sfregions}, along with Equation \ref{eq_sfr}. For
regions smaller than the beam areas accounting for uncertainties (i.e.
unresolved), $\Sigma_{\mathrm{SFR}}$ calculated here are lower-limits. To compare the
region size with values from the literature, we compute and report the effective
radius $R_e=\sqrt{\mathrm{area}/\pi}$, which has a range of 16 --
184\,pc for the 22 NRSF regions in the normal galaxies and 13 --
221\,pc for the 35 NRSF regions in the LIRGs. Regions in the normal
galaxies have SFR of 0.01 -- 0.21\,M$_\odot$yr$^{-1}$, with a median of
0.04$\pm$0.03\,M$_\odot$yr$^{-1}$. Regions in the LIRGs have SFR of
0.08 -- 1.7 \,M$_\odot$yr$^{-1}$, with a median of
0.25$\pm$0.12\,M$_\odot$yr$^{-1}$. Consequently, $\Sigma_{\mathrm{SFR}}$ for
regions in the LIRGs ranges from 7 -- 402\,M$_\odot$yr$^{-1}$kpc$^{-2}$ with a
median of 197$\pm$78\,M$_\odot$yr$^{-1}$kpc$^{-2}$, about an order of
magnitude higher compared to regions in the normal galaxies, whose
$\Sigma_{\mathrm{SFR}}$ ranges from 0.4 --
9.2\,M$_\odot$yr$^{-1}$kpc$^{-2}$ with a median of
1.4$\pm$0.9\,M$_\odot$yr$^{-1}$kpc$^{-2}$. We discuss the potential
effect of resolution on the results in Section \ref{dis:reg_sfr}. 

\subsection{Thermal Fractions at 33 GHz}\label{sec:regionfth} 
\begin{figure*}[t!]
    \epsscale{0.9}
    \plotone{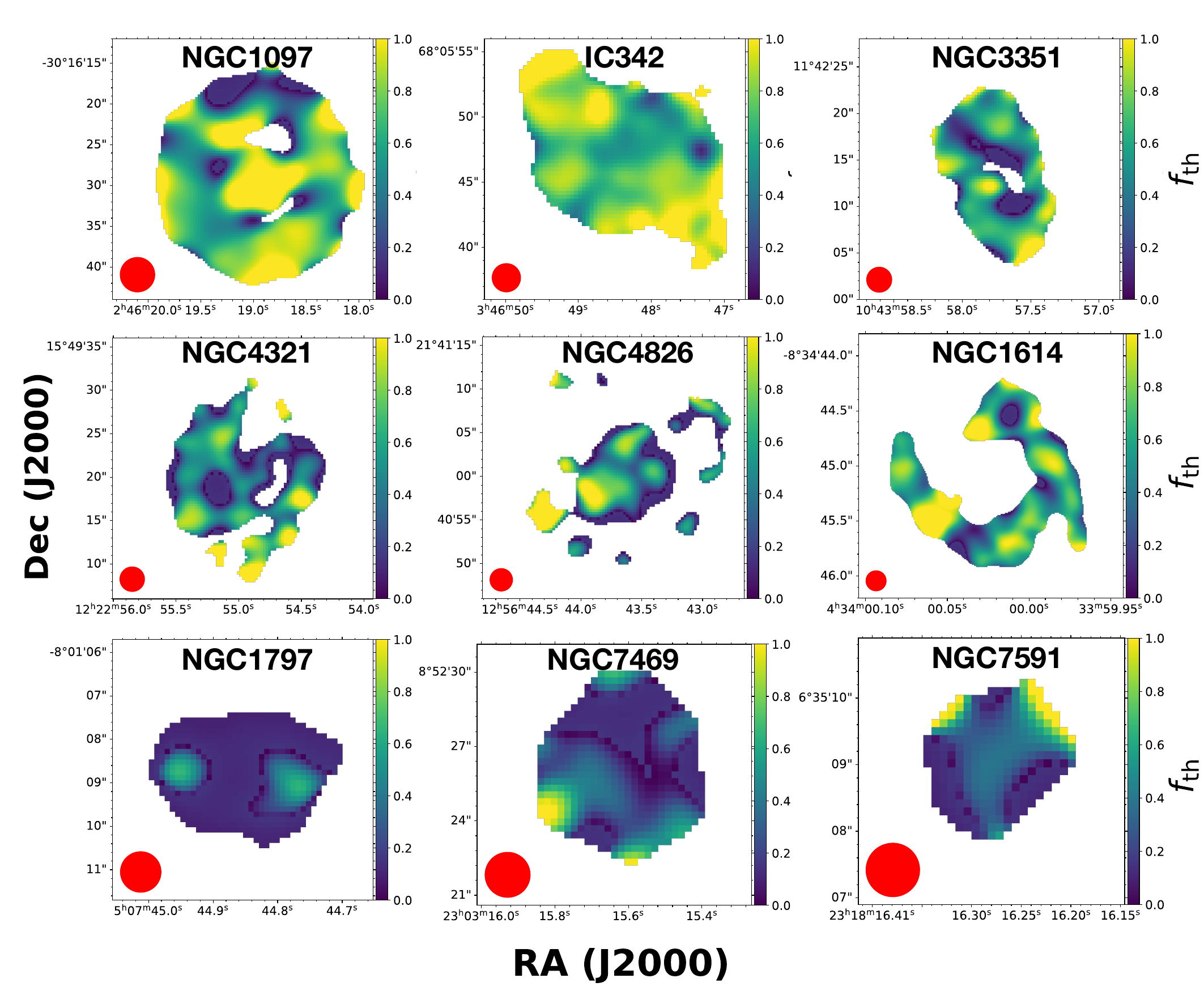}
    \caption{Maps of thermal fractions at 33 GHz of the sample galaxies. Each
    map was calculated from a pair of beam-matched 15 and 33 GHz images
    as described in Section \ref{sec:regionfth}. Red filled circles
    on the lower left of the maps represent the final matched beam. These maps
    show significant spatial variation in the distribution of thermal emission
    in these nuclear rings. Lower resolution images are used for NGC 1797, NGC
    7469 and NGC 7591. \label{fig:dendro_fth}}
\end{figure*}

Radio
continuum captures both non-thermal synchrotron emission of cosmic ray electrons
accelerated by supernovae, and thermal free-free emission associated with HII
regions of massive stars ($<$ 10\,Myr). At high radio frequencies, radio
emission has been directly related to ionizing photon produced by young massive
stars \citep{murphy12}, with thermal fractions (i.e. ratio of thermal free-free
emission to total radio continuum emission) $\gtrsim 90$\% at 33 GHz in
individual star-forming regions in nearby spiral galaxies \citep{linden20}.
However, it has been shown that even at 33 GHz, radio continuum emission in
local U/LIRGs may be largely non-thermal, due to dust absorption of ionizing
photons \citep{barcos15,barcos17}.\\
\indent To understand what is driving the 33 GHz radio continuum
emission in our sample of nuclear rings on sub-kpc scales, we estimate thermal fractions at 33
GHz, $f_\mathrm{th}$, using spectral index measured between 15 and 33 GHz (see
Section \ref{sec:sfregions}), and Equation (11) from \cite{murphy12}:
\begin{equation}\label{eq_fth}
    f^{\nu_1}_\mathrm{th}=\dfrac{\Big(\frac{\nu_2}{\nu_1}\Big)^{\alpha} - \Big(\frac{\nu_2}{\nu_1}\Big)^{\alpha_{\mathrm{NT}}}}{\Big(\frac{\nu_2}{\nu_1}\Big)^{-0.1} - \Big(\frac{\nu_2}{\nu_1}\Big)^{\alpha_{\mathrm{NT}}}}
\end{equation}
where $\alpha$ is the spectral index measured between $\nu_1$ and $\nu_2$ (33
and 15 GHz), and $\alpha_{\mathrm{NT}}$ is the non-thermal spectral index.
Almost all regions have $-0.85 < \alpha < -0.1$ within uncertainties, therefore
we adopt $\alpha_{\mathrm{NT}}=-0.85$ for our calculations following
\cite{murphy12}. Out of the total 58 NRSF regions, three have $\alpha \lesssim
-0.85$, which means their radio continuum spectra between 15 and 33 GHz are
steeper than the typical non-thermal spectrum. In these cases we set
$\alpha_{\mathrm{NT}}=\alpha-0.1$ based on previous measurements of the maximum
dispersion of non-thermal spectral index given by
$\sigma_{\alpha_{\mathrm{NT}}}\simeq0.1$ \citep{niklas97,murphy11,murphy12}.
Additionally, three NRSF regions have $\alpha \gtrsim -0.1$, in which cases we
set $f_\mathrm{th}$ to 100\%, assuming the spectral flattening is caused by
increasing thermal fraction. Alternatively, it may have originated from
$\alpha_{\mathrm{NT}}$ flattening or anomalous microwave emission
\citep[e.g.][]{dickinson18}, which will require future
matched-resolution observations at more than two radio frequencies to confirm.
\\
\indent Here we use only 15 and 33 GHz images to estimate $f_\mathrm{th}$
because 3 GHz observations from the GOALS equatorial survey do not resolve the
ring structures except for in NGC 7469. \cite{linden20} observed similar
spectral steepness at 3 -- 33 GHz and 15 -- 33 GHz for the full SFRS sample,
therefore we do not expect to overestimate 33 GHz thermal fractions in NRSF
regions in the normal galaxies by using only 15 and 33 GHz measurements.
However, in a study of extra-nuclear SF regions in local LIRGs, \cite{linden19}
observed steeper spectral profile at 3 -- 33 GHz compared to 15 -- 33 GHz on kpc
scales. Therefore the thermal fractions estimated at 33 GHz for the
sample of NRSF regions in LIRGs may be higher than reported here when
matched-resolution observations at 3 GHz are included, despite that we did not
observe significant spectral flattening at 15 -- 33 GHz compared to 3 -- 33 GHz
in NGC 7469.\\
\indent To better visualize the free-free emission distribution in these rings,
we additionally construct pixel-by-pixel maps of $f_\mathrm{th}$, as shown in
Figure \ref{fig:dendro_fth}. We mask all values below 5$\sigma_\mathrm{rms}$ for
15 and 33 GHz beam-matched images to ensure reliable outputs, after which we
calculate $\alpha$ and $f_\mathrm{th}$ at each pixel. These maps show
significant spatial variation in the distribution of thermal emission in these
nuclear rings, with areas of high $f_\mathrm{th}$ mostly corresponding to the
identified NRSF regions. Variations in $f_\mathrm{th}$ on scales smaller than
the matched beams are highly correlated and therefore not physically
significant.\\
\indent The estimated $f_\mathrm{th}$ has a range of 35 -- 100\% and 4 -- 100\%
for NRSF regions in the normal galaxies and in the LIRGs, respectively. We note
that low resolution ($\sim$ 0\farcs6 or 200pc) 15 and 33 GHz images were used to
calculate $f_\mathrm{th}$ for NGC 1797, NGC 7469 and NGC 7591 because the high
resolution images do not have strong enough detection for robust measurements.
This means that the physical scales at which $f_\mathrm{th}$ are measured are
2-5 times larger than the spatial extent of the identified regions,
likely including areas with little SF or diffuse non-thermal emission
(i.e. cosmic rays accelerated by supernovae). This can skew $f_\mathrm{th}$
towards lower values, and therefore values reported in Table \ref{tab:resolved}
may be interpreted as lower-limits. We further discuss the implications of these
results in Section \ref{dis:reg_fth}.

\startlongtable 
\centerwidetable
\begin{deluxetable*}{ccccrrcrrc}
\tablecaption{Region Properties
\label{tab:resolved}} \setcounter{table}{5} \tablecolumns{10}
\tablewidth{0pt} \tablehead{\colhead{Galaxy} & \colhead{$\nu$(GHz)} &
\colhead{ID} & \colhead{S$_\nu$(mJy)} & \colhead{$R_e$(pc)} &
\colhead{SFR(M$_\odot$ yr$^{-1}$)} &
\colhead{$\Sigma_\mathrm{SFR}$(M$_\odot$yr$^{-1}$kpc$^{-2}$)} &
\colhead{$\alpha$} & \colhead{$f_\mathrm{th}$(\%)}&
\colhead{Nucleus}\\
\colhead{(1)} & \colhead{(2)} & \colhead{(3)} & \colhead{(4)} &
\colhead{(5)} & \colhead{(6)} & \colhead{(7)} & \colhead{(8)} &
\colhead{(9)} & \colhead{(10)}
}
\startdata 
NGC 1097 & 33 & 1  & 0.92$\pm$0.12  & 142$\pm$22    & 0.10$\pm$0.01      & 1.5$\pm$0.1     & -0.18$\pm$0.23 & 91$\pm$24   &           \\
..       & 33 & 2  & 0.85$\pm$0.16  & 117$\pm$25    & 0.09$\pm$0.02      & 2.0$\pm$0.1     & -0.25$\pm$0.22 & 84$\pm$24   &           \\
..       & 33 & 3  & 2.00$\pm$0.25  & 184$\pm$27    & 0.21$\pm$0.03      & 1.9$\pm$0.1     & -0.44$\pm$0.20 & 62$\pm$26   &           \\
..       & 33 & 4  & 1.09$\pm$0.08  & 123$\pm$12    & 0.11$\pm$0.01      & 2.4$\pm$0.1     & -0.50$\pm$0.21 & 54$\pm$28   &           \\
..       & 33 & 5  & 3.48$\pm$0.04  & 163$\pm$6\;\; & N/A \;\;               & N/A         & -0.02$\pm$0.19 & 100$\pm$17  & AGN         \\
..       & 33 & 6  & 0.64$\pm$0.11  & 113$\pm$21    & 0.07$\pm$0.01      & 1.6$\pm$0.1     & -0.52$\pm$0.23 & 51$\pm$32   &           \\
..       & 33 & 7  & 0.89$\pm$0.60  & 121$\pm$85    & 0.09$\pm$0.06      & 2.0$\pm$0.5     & -0.32$\pm$0.23 & 77$\pm$27   &           \\
..       & 33 & 8  & 0.42$\pm$0.07  & 104$\pm$18    & 0.04$\pm$0.01      & 1.3$\pm$0.1     & -0.05$\pm$0.30 & 100$\pm$28  &           \\
..       & 33 & 9  & 1.40$\pm$0.08  & 164$\pm$14    & 0.15$\pm$0.01      & 1.7$\pm$0.1     & -0.40$\pm$0.21 & 66$\pm$26   &           \\
IC 342   & 33 & 1  & 12.3$\pm$0.42  & 49$\pm$3\;\;  & 0.07$\pm$0.01      & 9.2$\pm$0.2     & -0.33$\pm$0.18 & 75$\pm$21   &           \\
..       & 33 & 2  & 0.86$\pm$0.12  & 16$\pm$3\;\;  & 0.01$\pm$0.01      & 5.9$\pm$0.3     & -0.36$\pm$0.19 & 72$\pm$23   &           \\
..       & 33 & 3  & 2.10$\pm$0.25  & 23$\pm$3\;\;  & 0.01$\pm$0.01      & 6.8$\pm$0.3     & -0.27$\pm$0.18 & 82$\pm$20   &           \\
..       & 33 & 4  & 0.86$\pm$0.12  & 17$\pm$3\;\;  & 0.01$\pm$0.01      & 5.4$\pm$0.3     & -0.20$\pm$0.19 & 90$\pm$20   &           \\
NGC 3351 & 33 & 1  & 1.50$\pm$0.07  & 167$\pm$15    & 0.07$\pm$0.01      & 0.78$\pm$0.02   & -0.44$\pm$0.19 & 62$\pm$24   &           \\
..       & 33 & 2  & 0.19$\pm$0.04  & 58$\pm$13     & 0.01$\pm$0.01      & 0.82$\pm$0.06   & -0.62$\pm$0.22 & 37$\pm$33   &           \\
..       & 33 & 3  & 1.20$\pm$0.05  & 118$\pm$8\;\; & 0.05$\pm$0.01      & 1.2$\pm$0.1     & -0.43$\pm$0.18 & 62$\pm$24   &           \\
NGC 4321 & 33 & 1  & 0.06$\pm$0.01  & $<$72\;\;     & 0.01$\pm$0.01      &\;$>$0.42        & -0.30$\pm$0.35 & 79$\pm$40   &           \\
..       & 33 & 2  & 0.33$\pm$0.03  & 119$\pm$15    & 0.04$\pm$0.01      & 0.79$\pm$0.04   & -0.35$\pm$0.21 & 73$\pm$25   &           \\
..       & 33 & 3  & 0.20$\pm$0.04  & 110$\pm$26    &  N/A \;\;             &  N/A         & -0.63$\pm$0.24 & 35$\pm$36   & AGN     \\
..       & 33 & 4  & 0.17$\pm$0.03  & 85$\pm$17     & 0.02$\pm$0.01      & 0.80$\pm$0.06   & -0.52$\pm$0.22 & 51$\pm$31   &           \\
..       & 33 & 5  & 0.09$\pm$0.03  & 77$\pm$25     & 0.01$\pm$0.01      & 0.51$\pm$0.06   & -0.61$\pm$0.29 & 38$\pm$42   &           \\
..       & 33 & 6  & 0.11$\pm$0.03  & 68$\pm$19     & 0.01$\pm$0.01      & 0.78$\pm$0.07   & -0.47$\pm$0.23 & 57$\pm$31   &           \\
NGC 4826 & 33 & 1  & 0.47$\pm$0.05  & 44$\pm$6\;\;  &  N/A \;\;             &  N/A         & -0.47$\pm$0.19 & 58$\pm$26   & AGN      \\
..       & 33 & 2  & 0.44$\pm$0.03  & 40$\pm$4\;\;  & 0.01$\pm$0.01      & 1.3$\pm$0.1     & -0.51$\pm$0.19 & 53$\pm$26   &           \\
..       & 33 & 3  & 0.34$\pm$0.03  & 38$\pm$4\;\;  & 0.01$\pm$0.01      & 1.1$\pm$0.1     & -0.30$\pm$0.21 & 78$\pm$24   &           \\
NGC 1614 & 33 & 1  & 0.27$\pm$0.03  & 27$\pm$4\;\;  & 0.65$\pm$0.07      & 275$\pm$47\;\:  & -0.55$\pm$0.31 & 47$\pm$44   &           \\
..       & 33 & 2  & 0.06$\pm$0.01  & 15$\pm$3\;\;  & 0.14$\pm$0.03      & 213$\pm$68\;\;  & -0.38$\pm$0.39 & 68$\pm$48   &           \\
..       & 33 & 3  & 0.05$\pm$0.04  & 14$\pm$9\;\;  & 0.13$\pm$0.08      & 208$\pm$193     & -0.31$\pm$0.38 & 77$\pm$44   &           \\
..       & 33 & 4  & 0.04$\pm$0.01  & $<$14\;\;     & 0.10$\pm$0.03      & $>$149          & -0.79$\pm$0.36 & 10$\pm$60   &           \\
..       & 33 & 5  & 0.08$\pm$0.02  & 18$\pm$5\;\;  & 0.20$\pm$0.06      & 204$\pm$83\;\;  & -0.28$\pm$0.44 & 81$\pm$50   &           \\
..       & 33 & 6  & 0.17$\pm$0.19  & 22$\pm$26     & 0.41$\pm$0.45      & 266$\pm$433     & -0.07$\pm$0.30 & 100$\pm$29  &           \\
..       & 33 & 7  & 0.08$\pm$0.02  & 16$\pm$5\;\;  & 0.19$\pm$0.06      & 241$\pm$110     & -0.33$\pm$0.34 & 76$\pm$39   &           \\
..       & 33 & 8  & 0.08$\pm$0.02  & 16$\pm$4\;\;  & 0.19$\pm$0.04      & 249$\pm$88\;\;  & -0.43$\pm$0.32 & 63$\pm$40   &           \\
..       & 33 & 9  & 0.03$\pm$0.02  & $<$14\;\;     & 0.08$\pm$0.03      & $>$136          & -0.74$\pm$0.34 & 18$\pm$55   &           \\
..       & 33 & 10 & 0.10$\pm$0.06  & 16$\pm$10     & 0.25$\pm$0.15      & 306$\pm$277     & -0.39$\pm$0.33 & 67$\pm$41   &           \\
..       & 33 & 11 & 0.04$\pm$0.01  & $<$14\;\;     & 0.09$\pm$0.03      & $>$135          & -0.85$\pm$0.57 & 15$\pm$86   & Starburst \\
..       & 33 & 12 & 0.06$\pm$0.01  & $<$14\;\;     & 0.15$\pm$0.03      & $>$241          & -0.46$\pm$0.29 & 59$\pm$38   &           \\
..       & 33 & 13 & 0.20$\pm$0.03  & 22$\pm$4\;\;  & 0.48$\pm$0.08      & 311$\pm$82\;\;  & -0.27$\pm$0.30 & 82$\pm$33   &           \\
..       & 33 & 14 & 0.15$\pm$0.03  & 18$\pm$6\;\;  & 0.36$\pm$0.07      & 327$\pm$94\;\;  & -0.36$\pm$0.28 & 71$\pm$34   &           \\
..       & 33 & 15 & 0.45$\pm$0.07  & 31$\pm$4\;\;  & 1.1$\pm$0.16       & 366$\pm$85\;\;  & -0.26$\pm$0.25 & 83$\pm$27   &           \\
..       & 33 & 16 & 0.12$\pm$0.02  & 19$\pm$3\;\;  & 0.28$\pm$0.05      & 236$\pm$66\;\;  & -0.45$\pm$0.40 & 60$\pm$52   &           \\
..       & 33 & 17 & 0.04$\pm$0.01  & 13$\pm$4\;\;  & 0.10$\pm$0.03      & 197$\pm$83\;\;  & -0.03$\pm$0.43 & 100$\pm$40  &           \\
..       & 33 & 18 & 0.07$\pm$0.02  & 17$\pm$6\;\;  & 0.17$\pm$0.06      & 207$\pm$103     & -0.36$\pm$0.33 & 71$\pm$40   &           \\
..       & 33 & 19 & 0.10$\pm$0.03  & 14$\pm$4\;\;  & 0.25$\pm$0.06      & 402$\pm$150     & -0.89$\pm$0.24 & 15$\pm$34   &           \\
..       & 33 & 20 & 0.09$\pm$0.02  & $<$14\;\;     & 0.21$\pm$0.04      & $>$326          & -0.80$\pm$0.24 & 8$\pm$41    &           \\
..       & 33 & 21 & 0.16$\pm$0.03  & 19$\pm$4\;\;  & 0.38$\pm$0.07      & 316$\pm$92\;\;  & -0.55$\pm$0.27 & 47$\pm$38   &           \\
NGC 1797 & 33 & 1  & 0.18$\pm$0.02  & $<$100        & 0.38$\pm$0.05      & $>$12\;           & -0.78$\pm$0.21 & 11$\pm$35   &           \\
..       & 33 & 2  & 0.68$\pm$0.03  & 194$\pm$12    & 1.4$\pm$0.05       & 12$\pm$1\;\;    & -0.67$\pm$0.19 & 28$\pm$30   &           \\
..       & 33 & 3  & 0.35$\pm$0.04  & 126$\pm$17    & 0.73$\pm$0.08      & 14$\pm$3\;\;    & -0.68$\pm$0.20 & 28$\pm$32  &           \\
..       & 33 & 4  & 0.11$\pm$0.02  & $<$99\;\;     & 0.23$\pm$0.04      & $>$7.4          & -1.19$\pm$0.22 & 12$\pm$26   &           \\
NGC 7469 & 33 & 1  & 0.64$\pm$0.11  & 221$\pm$45    & 1.7$\pm$0.29       & 11$\pm$3\;\;    & -0.57$\pm$0.18 & 43$\pm$27   &           \\
..       & 33 & 2  & 0.60$\pm$0.07  & 157$\pm$24    & 1.6$\pm$0.19       & 20$\pm$4\;\;    & -0.79$\pm$0.18 & 10$\pm$31   &           \\
..       & 33 & 3  & 6.06$\pm$0.20  & 268$\pm$30    &  N/A  \;\;             &  N/A        & -0.74$\pm$0.18 & 19$\pm$29   & AGN       \\
..       & 33 & 4  & 0.17$\pm$0.04  & 90$\pm$23     & 0.45$\pm$0.11      & 18$\pm$6\;\;    & -0.83$\pm$0.18 & 4$\pm$32    &           \\
..       & 33 & 5  & 0.24$\pm$0.04  & 96$\pm$20     & 0.61$\pm$0.11      & 21$\pm$6\;\;    & -0.74$\pm$0.18 & 17$\pm$30   &           \\
..       & 33 & 6  & 0.19$\pm$0.03  & $<$95\;\;     & 0.49$\pm$0.07      & \;$>$17           & -0.88$\pm$0.18 & 15$\pm$27   &           \\
NGC 7591 & 15 & 1  & 0.14$\pm$0.03  & 21$\pm$5\;\:  & 0.27$\pm$0.07      & 197$\pm$69\;\;  & -0.64$\pm$0.18 & 34$\pm$27   &           \\
..       & 15 & 2  & 0.12$\pm$0.04  & 21$\pm$7\;\:  & 0.22$\pm$0.07      & 165$\pm$77\;\;  & -0.66$\pm$0.18 & 30$\pm$28   &           \\
..       & 15 & 3  & 0.12$\pm$0.02  & 19$\pm$5\;\:  & 0.22$\pm$0.05      & 197$\pm$63\;\;  & -0.63$\pm$0.18 & 35$\pm$27   &           \\
..       & 15 & 4  & 0.09$\pm$0.02  & 19$\pm$5\;\:  & 0.17$\pm$0.04      & 139$\pm$46\;\;  & -0.65$\pm$0.18 & 32$\pm$28   &           \\
..       & 15 & 5  & 0.05$\pm$0.01  & 18$\pm$5\;\:  & N/A \;\;           &  N/A           & -0.63$\pm$0.18 & 35$\pm$27   & AGN       \\
..       & 15 & 6  & 0.11$\pm$0.02  & 22$\pm$6\;\:  & 0.21$\pm$0.04      & 130$\pm$43\;\;  & -0.63$\pm$0.18 & 35$\pm$27   &           \\
..       & 15 & 7  & 0.40$\pm$0.24  & 46$\pm$30     & 0.76$\pm$0.47      & 115$\pm$103     & -0.64$\pm$0.18 & 34$\pm$28   &          
\enddata
\tablecomments{(1): Host galaxy of the nuclear ring; (2): Frequency at which
        regions were identified and SFR and sizes were measured. For NGC 7591,
        15GHz image was used instead. (3): Identifier of the region in reference
        to Figure \ref{fig:dendro_sfr}. (4): Flux density of the region. (5):
        Effective radius of the region assuming region is circular (Section
        \ref{sec:astrodendro}). Beam areas are used for unresolved regions as
        upper-limits for their sizes, indicated by ``$<$".  (6): SFR calculated
        using Eq.\ref{eq_sfr} and (5) for non-AGN regions; (7):
        $\Sigma_\mathrm{SFR}$ estimated by dividing (6) over region area $A=\pi
        R^2_e$ (in kpc$^2$). Unresolved regions are given lower-limits for their
        $\Sigma_\mathrm{SFR}$, indicated by ``$>$"; (8): Spectral index
        associated with the region (Section \ref{sec:regionfth}) measured from
        15 to 33 GHz. (9): Fraction of thermal emission at 33 GHz estimated
        using $\alpha$ (Section \ref{sec:sfregions}). Given the coarser
        resolution of the beam-matched 15 and 33 GHz images on which $\alpha$
        were measured, the reported 33 GHz thermal fractions may be considered
        as lower-limits. See Section \ref{dis:reg_fth} for discussion. (10): Whether the region corresponds to a galactic
        nucleus (AGN/starburst), see Table \ref{tab:sample} for references.}
\end{deluxetable*}

\subsection{Gas depletion times}\label{sec:tdep_result}
\begin{figure*}[h!]
\epsscale{1.2}
\plotone{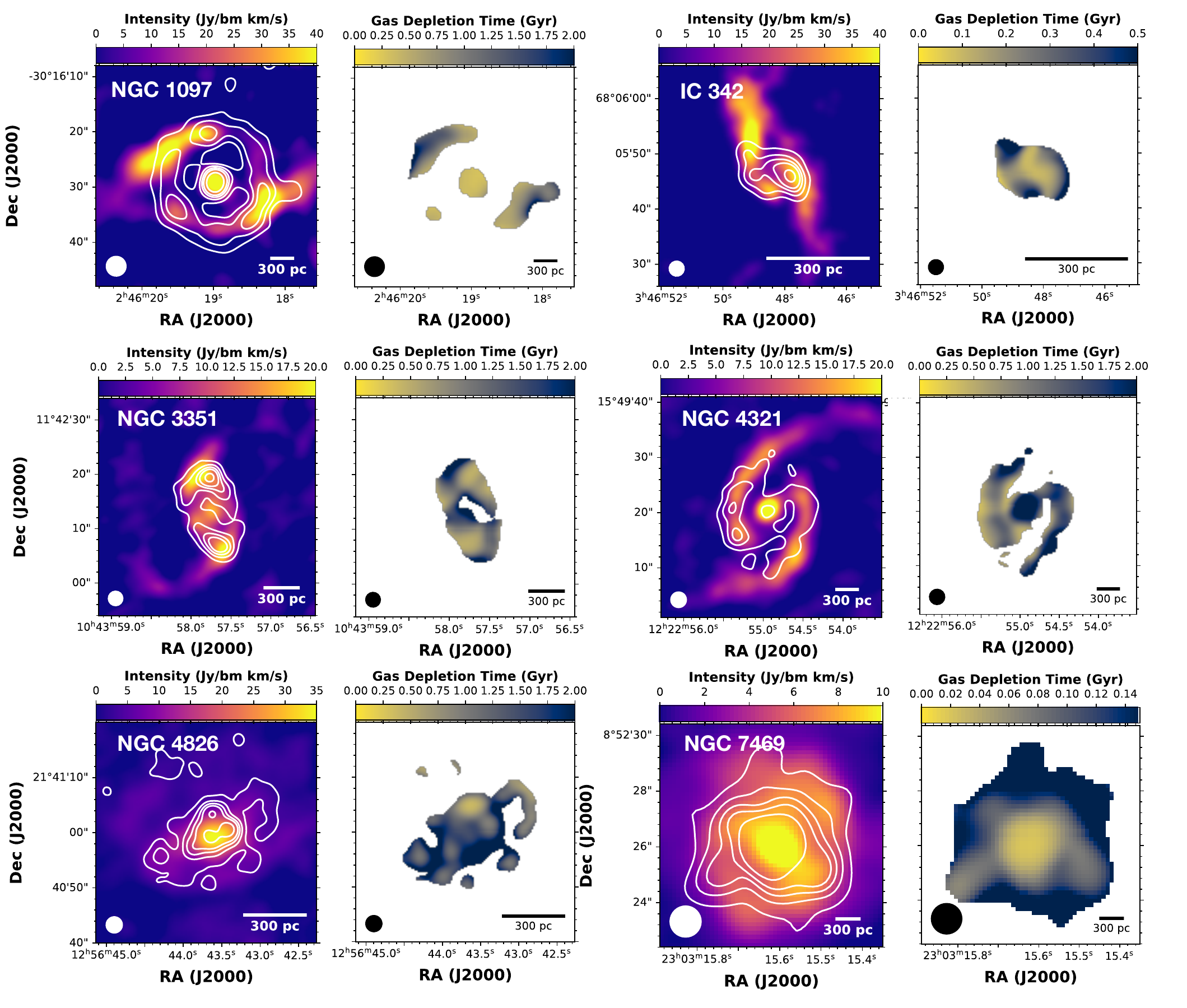}
\caption{Six nuclear rings with high-resolution archival CO (J=1-0) datasets. For each galaxy panel, \textit{left:}
Beam-matched archival CO(J=1-0) moment 0 map (in color) and VLA 33 GHz continuum
image (white contour). The contours corresponds to $5, 10, 15, 20,
30\,\sigma_\mathrm{rms}$ levels in 33 GHz intensity; \textit{right:} Maps of gas
depletion time $\tau_\mathrm{dep} = \Sigma_\mathrm{mol}/\Sigma_\mathrm{SFR}$,
using maps shown on the \textit{left}. Matched beams are represented by
white/black filled circles on the lower left of each image in the left/right
panel. The radio continuum is largely co-spatial with cold molecular
gas. In NGC 1097, IC 342, NGC 3351, and NGC 4321, prominent spiral structures
overlap with the SF nuclear rings and extend farther out into the galactic
disks. The spiral structure in NGC 7469 is much more tightly wound and less
distinct, and its nuclear ring has an order of magnitude shorter gas depletion
times on average.}\label{fig:ka_co}
\end{figure*}

\indent
In the left panels of Figure \ref{fig:ka_co}, we show the beam-matched CO
(J=1-0) moment 0 maps (in color) and 33 GHz continuum data (in contour) for the
six nuclear rings in the sample that have archival CO (J=1-0) data at resolutions
comparable to the VLA data (Section \ref{sec:almadata}). We can see that the nuclear rings
observed in the radio continuum are largely co-spatial with the cold molecular
gas, and molecular spiral arms are visible beyond the rings in NGC
1097, IC 342, NGC 3351 and NGC 4321. Using the measurements of CO (J=1-0) and 33
GHz continuum emission on these resolution-matched maps, we can calculate the
cold molecular gas mass ($M_\mathrm{mol}$) and surface densities
($\Sigma_\mathrm{mol}$) in these six nuclear rings and their individual NRSF
regions, and make direct comparisons with the SFR and $\Sigma_\mathrm{SFR}$ to
estimate the timescale at which SF depletes the molecular gas, which is used in
both observational and theoretical studies to quantify star formation
efficiencies \citep[i.e. $\tau_\mathrm{dep}$ = 1/SFE =
$\Sigma_\mathrm{mol}/\Sigma_\mathrm{SFR}$; e.g.][]{bigiel08,wilson19,moreno21}.
\\
\indent We follow \cite{herrero-illana19} and use equation
from \cite{solomon92} to convert the measured CO(J=1-0) flux to
molecular gas mass: 
\begin{align}\label{eq_gas}
    M_\mathrm{mol}&=\alpha_\mathrm{CO} L'_\mathrm{CO}\\
    &=2.45\times10^3\alpha_\mathrm{CO}\Big(\frac{S_\mathrm{CO}\Delta\nu}{\mathrm{Jy\ km/s}}\Big)\Big(\frac{D_L}{\mathrm{Mpc}}\Big)^2(1+z)^{-1}
\end{align}
where $M_\mathrm{mol}$ is in units of M$_\odot$,
$S_\mathrm{CO}\Delta\nu$ is the integrated line flux in Jy km/s and
$D_L$ is the luminosity distance in Mpc reported in Table
\ref{tab:sample} given the redshift $z$. Finally, $\alpha_\mathrm{CO}$ is the
CO-to-H$_2$ conversion factor, in units of M$_\odot$(K km s$^{-1}$
pc$^{-2}$)$^{-1}$.  We adopt $\alpha_\mathrm{CO}=$ 4.3 M$_\odot$(K km s$^{-1}$
pc$^{-2}$)$^{-1}$ for the five nuclear rings hosted in normal galaxies following
previous resolved studies of nearby disk galaxies \citep{bigiel08,leroy13}, and
we use the U/LIRG value $\alpha_\mathrm{CO}$ = 1.8 M$_\odot$(K km s$^{-1}$
pc$^{-2}$)$^{-1}$ from \cite{herrero-illana19} for the nuclear ring in NGC 7469. We
use Equation \ref{eq_sfr} to convert the 33 GHz continuum flux to SFR, and
$\Sigma_\mathrm{SFR}$ and $\Sigma_\mathrm{mol}$ are calculated using the
physical areas of the adopted apertures (See Section \ref{sec:tdep_analysis}).
In Table \ref{tab:gasdata} we summarize the derived quantities for the nuclear rings
and the individual NRSF regions. \\
\indent Based on calculations using previous global CO (J=1-0)
measurements \citep{young96,cros01,crosthwaite01, garcia03,davies04}, these six
nuclear rings contain $\sim$ 10 -- 30\% of the total molecular gas mass of their
host galaxies; this gas is available to fuel the active starbursts that are
responsible for $\sim$ 10 -- 60\% of the total SFR of the host galaxies. The
average $\Sigma_\mathrm{mol}$ has a range of $280\pm40$ --
$900\pm90$\,M$_\odot$yr$^{-1}$pc$^{-2}$ in these nuclear rings, and $84\pm180$
-- $1970\pm200$\,M$_\odot$yr$^{-1}$pc$^{-2}$ in the NRSF regions. The gas
depletion times $\tau_\mathrm{dep}$ associated with the individual NRSF regions
range from 0.07 -- 1.4\,Gyr. The median $\tau_\mathrm{dep}$ for regions in the
normal galaxies is 0.6$\pm$0.5\,Gyr. This is almost an order of magnitude longer
than regions in NGC 7469, which has a median $\tau_\mathrm{dep}$ of
0.08$\pm$0.01\,Gyr. These values agree with results from previous sub-kpc
studies of normal galaxies \citep{bigiel08,leroy13} and U/LIRGs
\citep{wilson19}. Despite that our measurements are made on scales larger than
the region sizes measured using native-resolution radio maps, the median
$\tau_\mathrm{dep}$ for the NRSF regions is largely consistent with
$\tau_\mathrm{dep}$ measured over the entire ring for each galaxy, therefore
higher-resolution measurements may increase the scatter of $\tau_\mathrm{dep}$
estimated for these regions but will not significantly change the results. In
the right panels of Figure \ref{fig:ka_co}, we also show pixel-by-pixel map of
$\tau_\mathrm{dep}$ for each nuclear ring for more direct visualization. We note
that $\tau_\mathrm{dep}$ derived near AGN is not meaningful as the 33 GHz
emission is not associated solely with star formation, and hence we also do not
report $\Sigma_\mathrm{SFR}$ and $\tau_\mathrm{dep}$ in Table
\ref{tab:gasdata} for regions containing AGN. We further discuss these results
in the context of universal star formation relation \citep[e.g.][]{kennicutt98}
in Section \ref{sec:tdep}.

\begin{deluxetable*}{c|lcccccc}
\tablecaption{Quantities derived from resolution-matched 33 GHz and CO(J=1-0) maps}
\label{tab:gasdata} \setcounter{table}{6} \tablecolumns{8}
\tablewidth{0pt} \tablehead{\colhead{} & \colhead{Galaxy} & 
\colhead{ID} & \colhead{A$_\mathrm{ap}$($\arcsec^2$)} & \colhead{$M_\mathrm{mol}$($10^6$\,M$_\odot$)} & \colhead{$\Sigma_\mathrm{mol}$(M$_\odot$yr$^{-1}$pc$^{-2}$)} &
\colhead{$\Sigma_\mathrm{SFR}$(M$_\odot$yr$^{-1}$kpc$^{-2}$)} &
\colhead{$\tau_\mathrm{dep}$(Gyr)} \\
\colhead{} & \colhead{(1)} & \colhead{(2)} & \colhead{(3)} & \colhead{(4)} &
\colhead{(5)} & \colhead{(6)} 
}                                                       
\startdata 
\multirow{6}{*}{Ring} & NGC 1097 &  - & 577 &\;\;\;\,760$\pm$100    & 280$\pm$40 & 0.72$\pm$0.07 & 0.38$\pm$0.06\\
                      & IC 342   &  - & 213 &\;\,47$\pm$5   & 860$\pm$90 & 2.9$\pm$0.3   & 0.29$\pm$0.04\\
                      & NGC 3351 &  - & 170 &\;\,190$\pm$20     & 550$\pm$60 & 0.55$\pm$0.05 & 1.0$\pm$0.1\\
                      & NGC 4321 &  - & 312 &\;\,430$\pm$40     & 280$\pm$30 & 0.26$\pm$0.03 & 1.1$\pm$0.2\\
                      & NGC 4826 &  - & 82 &\;\,50$\pm$5   & 790$\pm$80 & 0.58$\pm$0.06 & 1.4$\pm$0.2\\
                      & NGC 7469 &  - & 25 &2500$\pm$1\;\: & 890$\pm$90 & 8.8$\pm$0.9   & 0.10$\pm$0.01\\ \hline
\multirow{30}{*}{Region} & NGC 1097 &1& 13.4&\;\;\;\,26$\pm$10  & 410$\pm$20 & 1.2$\pm$0.1   & 0.35$\pm$0.14\\
                      & ..       &  2 & 10.14&\;\;\;\,47$\pm$10  & \;\;970$\pm$200 & 1.6$\pm$0.2  & 0.61$\pm$0.15 \\
                      & ..       &  3 & 22.5&\;\;\;\,49$\pm$14  & \;\;450$\pm$130 & 1.5$\pm$0.2  & 0.32$\pm$0.09\\
                      & ..       &  4 & 10.1&\;\;\;\,61$\pm$11  & 1260$\pm$220 & 1.5$\pm$0.2 & 0.82$\pm$0.17\\
                      & ..       &  5 & 17.6&\;\;\;\,90$\pm$15  & 1070$\pm$170 & N/A & N/A\\
                      & ..       &  6 & 10.1&\;\;\;\,35$\pm$10  & \;\;730$\pm$190 & 1.2$\pm$0.1      & 0.63$\pm$0.18\\
                      & ..       &  7 & 10.1&\;\,14$\pm$9   & \;\;290$\pm$180 & 1.3$\pm$0.1      & 0.23$\pm$0.15\\
                      & ..       &  8 & 10.1&\;\;\;\,4.1$\pm$8.7    & \;\;\;\;84$\pm$180  & 0.68$\pm$0.10& 0.12$\pm$0.27\\
                      & ..       &  9 & 18.0&\;\;\;\,63$\pm$13  & \;\;740$\pm$150 & 1.3$\pm$0.1      & 0.58$\pm$0.14\\
                      & IC 342   &  1 & 29.4&\;\,11$\pm$3   & 1460$\pm$150 & 7.7$\pm$0.8     & 0.19$\pm$0.03\\
                      & ..       &  2 & 5.93&\;\;\;\,2.5$\pm$1.1  & 1610$\pm$160 & 4.6$\pm$0.5   & 0.35$\pm$0.05\\
                      & ..       &  3 & 6.80&\;\;\;\,1.7$\pm$1.2  & \;\;990$\pm$100 & 5.1$\pm$0.6    & 0.19$\pm$0.03\\
                      & ..       &  4 & 5.93&\;\;\;\,3.0$\pm$1.1  & 1970$\pm$200 & 4.0$\pm$0.5   & 0.49$\pm$0.08\\
                      & NGC 3351 &  1 & 43.0&\;\,54$\pm$6     & 620$\pm$70   & 0.66$\pm$0.07 & 0.94$\pm$0.14\\
                      & ..       &  2 & 5.72&\;\;\;\,7.8$\pm$1.1  & \;\;680$\pm$100   & 0.71$\pm$0.08& 0.95$\pm$0.17\\
                      & ..       &  3 & 21.4&\;\,34$\pm$4     & 780$\pm$90   & 1.0$\pm$0.1   & 0.75$\pm$0.11\\
                      & NGC 4321 &  1 & 6.31&\;\,20$\pm$2     & 660$\pm$80   & 0.29$\pm$0.05 & 2.30$\pm$0.46\\
                      & ..       &  2 & 9.26&\;\,26$\pm$3     & 580$\pm$70   & 0.61$\pm$0.07 & 0.94$\pm$0.15\\
                      & ..       &  3 & 7.86&\;\,49$\pm$5     & 1270$\pm$130 & N/A & N/A\\
                      & ..       &  4 & 6.31&\;\,14$\pm$2     & 440$\pm$60   & 0.62$\pm$0.07 & 0.70$\pm$0.13\\
                      & ..       &  5 & 6.31&\;\,16$\pm$2     & 530$\pm$70   & 0.38$\pm$0.05 & 1.4$\pm$0.3 \\
                      & ..       &  6 & 6.31&\;\,18$\pm$2     & 590$\pm$70   & 0.57$\pm$0.07 & 1.0$\pm$0.2\\
                      & NGC 4826 &  1 & 9.39&\;\;\;\,9.2$\pm$0.9  & 1450$\pm$150 & N/A & N/A\\
                      & ..       &  2 & 7.58&\;\;\;\,5.8$\pm$0.6  & 1140$\pm$120 & 1.0$\pm$0.1 & 1.1$\pm$0.2\\
                      & ..       &  3 & 6.94&\;\;\;\,1.9$\pm$0.3  & 420$\pm$60   & 0.74$\pm$0.07 & 0.57$\pm$0.09\\
                      & NGC 7469 &  1 & 1.40&\;\:280$\pm$30   & \;\;750$\pm$100  & 8.5$\pm$0.9 & 0.09$\pm$0.01\\
                      & ..       &  2 & 0.98&\;\:290$\pm$30   & 1130$\pm$130 & 12$\pm$1\;\, & 0.09$\pm$0.01\\
                      & ..       &  3 & 2.04&\;\:800$\pm$80   & 1480$\pm$150 & N/A & N/A\\
                      & ..       &  1 & 0.98&\;\:290$\pm$30  & 1110$\pm$130 & 17$\pm$2\;\, & 0.07$\pm$0.01\\
                      & ..       &  2 & 0.98&\;\:270$\pm$30  & 1060$\pm$120 & 12$\pm$1\;\, & 0.08$\pm$0.01\\
                      & ..       &  3 & 0.98&\;\:280$\pm$30  & 1090$\pm$130 & 13$\pm$1\;\, & 0.09$\pm$0.01\\
\enddata
\tablecomments{(1): Host galaxy of the nuclear ring. (2): Identifier of
the region in reference to Figure \ref{fig:dendro_sfr}. (3): Area of the
circular aperture used to measure 33 GHz and CO(J=1-0) flux of the nuclear
ring/NRSF region. For IC 342 and NGC 4826, beam areas are subtracted from the
ring areas defined by $R_\mathrm{out}$ to account for the central unresolved
cavities.(4): Molecular gas mass derived from CO (J=1-0) flux measurements (see
Section \ref{sec:tdep_result}). (5): Molecular gas surface density over the
physical areas of the adopted apertures. (6): Star formation rate surface
density over the physical areas of the adopted apertures for non-AGN regions.
(7) Gas depletion time calculated using (5) and (6) for non-AGN regions.}
\end{deluxetable*}

\section{Discussion}\label{sec:discuss}
Our selection criteria has resulted in the identification of NRSF in four LIRGs
and five normal galaxies. The NRSF in our sample exhibits diverse spatial
distributions (Figure \ref{fig:natres1} and \ref{fig:natres2}). For examples,
NGC 1097 and NGC 1614 have more randomly distributed NRSF regions along the
rings compared to NGC 3351 and NGC 1797, where bright regions occur on opposite
sides of the ring. Several studies have discussed the potential mechanisms that
may give rise to certain alignments of bright NRSF regions, such as orbit
crowding of gas clouds at the ends of nuclear stellar bars
\citep[e.g.,][]{kenney91,mazzarella94,englmaier04}, or specific gas inflow rates
into the ring \citep[e.g.,][]{seo13}. Depending on how gas accumulates, NRSF can either take place stochastically in the ring due to
gravitational instability, resulting in random spatial distribution of ``hot
spots'', or close to the contact points between the ring and dust lanes in
multiple bursts \citep[e.g.][]{boeker08}. Given the limited sample size, we do
not further discuss the implications associated with the NRSF spatial
distribution, but simply provide descriptions of our observations and relevant
information from previous studies on individual nuclear rings in Appendix A.
Instead, we focus our discussions on the implications of the results in
Section \ref{sec:results}, and the limitations of the present sample.
\subsection{Sample limitation}\label{dis:bias} The sample presented in
this work is limited by several effects. Given the resolution of the
observations, a nuclear ring has to have an angular radius larger than the
synthesized beam in order for the ring structure to be resolved. An example is
NGC 4579, a known nuclear ring host in SFRS, whose ring radius is estimated to
be 1.6\arcsec\ from HST optical and near-IR observations \citep{comeron10},
which is smaller than the 33 GHz beam size of $\sim$ 2\arcsec, and therefore not
included in our sample. Additionally, the surface brightness of the nuclear ring
must be high enough above the sensitivity limits of the observations for the
ring structure to be visually distinct. NGC 4736 and NGC 5194 are two other
galaxies from SFRS that are included in \cite{comeron10} but excluded from our
sample because many of their NRSF regions are too faint for the ring structure
to be visually distinct. This may be due to overall lower level of SF activity,
and/or a lack of sensitivity of the observations. For example, with an estimated
angular radius of 50\arcsec \citep[1.2\,kpc,][]{comeron10}, the ring in NGC 4736
is detected close to the edge of the 33 GHz primary beam, where sensitivity is
significantly worse compared to the phase center, which results in incomplete
detection of the ring structure.\\
\indent For observations from the GOALS equatorial survey, rings at very
close distances such as the one in NGC 1068 become too highly-resolved at A
configuration to be detected given the sensitivity limit and rings that are far
away may either have been unresolved or lack consistent detection of NSRF
regions for the ring structure to be visually identified. The fact that all four
LIRGs in our sample have similar luminosity distances at $\sim$ 70\,Mpc may be
the result of such a trade-off between physical resolution and sensitivity.
Finally, rings that are highly inclined may appear linear and therefore are not
represented in our sample. High-resolution kinematics studies are needed to
reveal these edge-on rings.\\
\indent Given the above, our sample represents a lower-limit of the
number of nuclear rings in both surveys, and thus the results derived in this
study may not represent the full range of NRSF properties in SFRS and the GOALS
equatorial survey.\\
\subsection{The majority of SF in these LIRGs takes place in their nuclear
rings}\label{dis:ringsfr} 
\begin{figure}[t!]
    \epsscale{1.2}
    \plotone{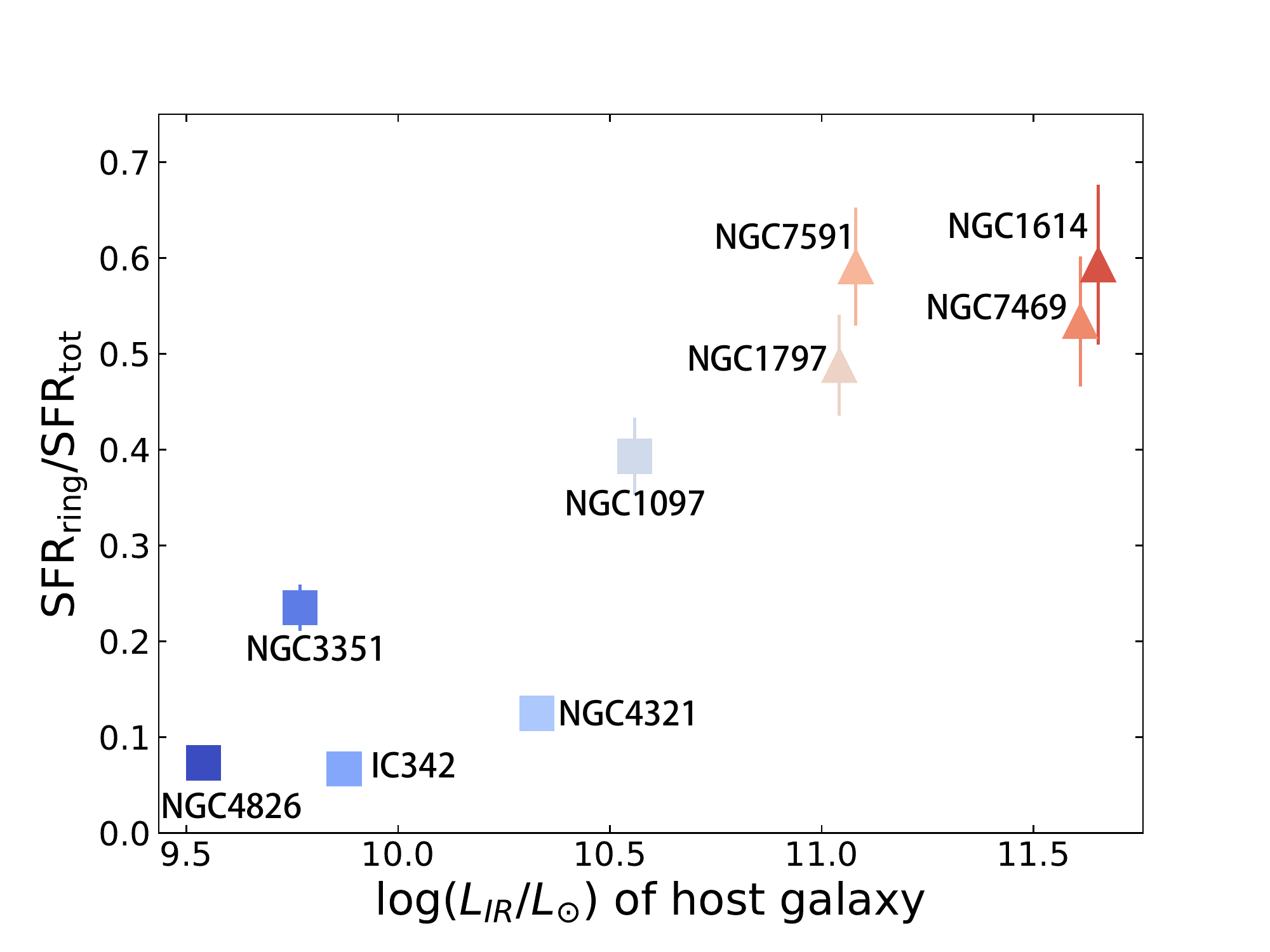} 
    \caption{Fraction of total SFR contributed by the nuclear ring with respect to
    $L_\mathrm{IR}$ of the host galaxy. See Section
    \ref{sec:integrated}, Table \ref{tab:integrated} and \ref{tab:host_sfr} for
    details. Each galaxy is color-coded by its $L_\mathrm{IR}$, with darker
    blue and darker red representing the lower and higher $L_\mathrm{IR}$,
    respectively. Squares and triangles represent normal galaxies and LIRGs
    respectively. \textbf{Nuclear rings in the LIRGs consistently contribute
    higher fractions of the total SFR of their host galaxies compared to
    rings in the normal galaxies.}\label{fig:ringfraction}}
\end{figure}

\indent Figure \ref{fig:ringfraction} shows the
fraction of total SFR contributed by the nuclear ring with respect to
$L_\mathrm{IR}$ of the host galaxy, as calculated in Section
\ref{sec:ring_frac}. Each galaxy is color-coded by its
$L_\mathrm{IR}$, with darker blue and darker red representing the lower and
higher $L_\mathrm{IR}$, respectively. Nuclear rings hosted in the LIRGs have up
to six times higher
$\frac{\mathrm{SFR}_\mathrm{ring}}{\mathrm{SFR}_\mathrm{tot}}$ than the ones
hosted in the normal galaxies. Furthermore, we can also see that high
$\frac{\mathrm{SFR}_\mathrm{ring}}{\mathrm{SFR}_\mathrm{tot}}$ in general
corresponds to galaxies with high $L_\mathrm{IR}$. This result echoes previous
studies which found that local galaxies with higher $L_\mathrm{IR}$ have more centrally
concentrated emission \citep{diaz-santos10,diaz-santos11}, and that the nuclear
SF in LIRGs can dominate the properties of their host galaxies
\citep[e.g.][]{veilleux95,soifer01}. However, the nuclear rings we study here
may only represent the most extreme cases, and it is possible that in
many LIRGs, the total star formation is less centrally concentrated. We will
present results on various nuclear SF structures in the entire GOALS equatorial
survey in a upcoming paper to further investigate this. \\ 
\indent It is also worth noting that NGC 1097, which is interacting with a dwarf
companion, has both the highest $L_\mathrm{IR}$ and the highest
$\frac{\mathrm{SFR}_\mathrm{ring}}{\mathrm{SFR}_\mathrm{tot}}$ among the normal
galaxies. This trend is consistent with studies that observed excess nuclear SFR
in interacting galaxies relative to isolated systems
\citep[e.g.][]{lonsdale84,bushouse86}, which is also predicted in simulations of
galaxy interaction \citep{moreno21}. 

\subsection{High SFR and $\Sigma_\mathrm{SFR}$ in NRSF regions in the sample of
LIRGs}\label{dis:reg_sfr} 
\begin{figure}[tbh!]
    \epsscale{1.2}
    \plotone{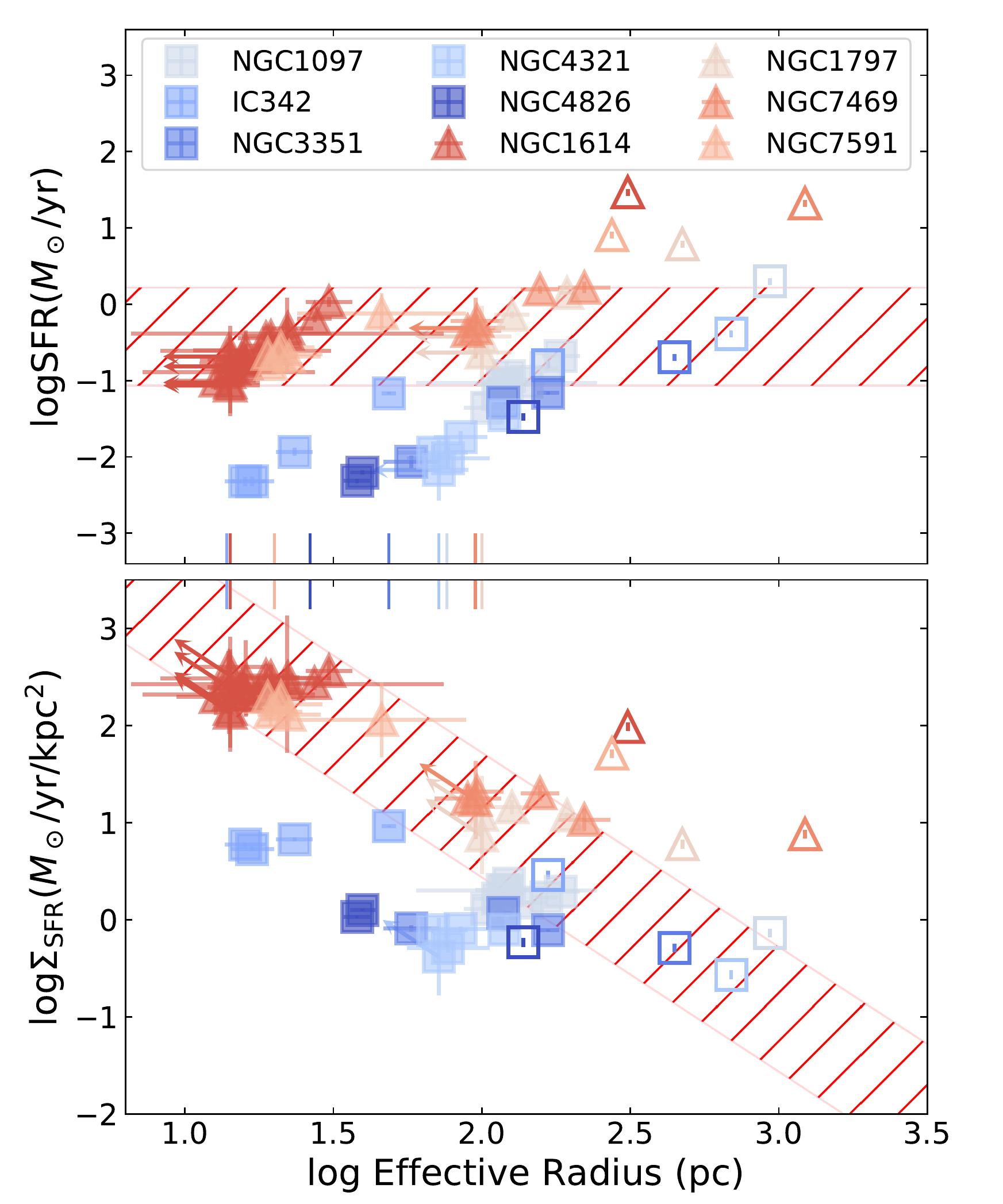} 
    \caption{Effective radius $R_e$ vs. SFR (\textit{top}) and
    $\Sigma_{\mathrm{SFR}}$ (\textit{bottom}). Values for the 58 NRSF regions
    are shown in filled symbols, and integrated ring values are in non-filled
    symbols, both color-coded by $L_\mathrm{IR}$ of the host galaxy, with
    squares and triangles representing normal galaxies and LIRGs, respectively.
    Arrows indicate upper-limits in $R_e$ for unresolved regions, which
    translate into lower-limits in $\Sigma_{\mathrm{SFR}}$ given the measured
    SFR. Short vertical lines on the horizontal axis mark the resolution
    limits of the observations. The red hatching represents the range of SFR and
    expected range of $\Sigma_{\mathrm{SFR}}$ spanned by the smallest and
    largest NRSF regions in the sample of LIRGs, assuming the extreme case where
    $\Sigma_\mathrm{SFR}$ is diluted by quiescent areas when measured within
    larger region size (i.e. $\Sigma_\mathrm{SFR} \propto 1/R^2_e$).
    \textbf{The NRSF regions studied in our sample of LIRGs have higher SFR and
    $\Sigma_\mathrm{SFR}$ with sizes similar to or smaller than their
    counterparts in the sample of normal galaxies. } \label{fig:sigsfr_re}}
\end{figure}

The consistently higher nuclear ring contribution to
the total SFR, as discussed in the last Section, points to more active NRSF in
our sample of LIRGs. In Figure \ref{fig:sigsfr_re}, we show, in filled symbols,
the SFR and $\Sigma_\mathrm{SFR}$ of all 58 NRSF regions with respect to their
effective radius $R_e$ (reported in Table \ref{tab:resolved}), and in non-filled
symbols, the integrated values for the entire rings (Table
\ref{tab:integrated}). Note that for the integrated values, $R_e =
\sqrt{\mathrm{area}/\pi}$ measures the effective area extended by the ring, and
is different from $R_\mathrm{peak}$ defined in Section \ref{sec:integrated}.\\
\indent Despite the different angular resolutions of the observations,
the physical resolutions achieved (marked in short vertical lines on the
horizontal axis) and the effective sizes of NRSF regions measured are similar between the
normal galaxies and the LIRGs. We can see that at similar or smaller effective
sizes, the integrated rings and NRSF regions in the LIRGs (triangles) both have
at least an order of magnitude higher SFR and $\Sigma_{\mathrm{SFR}}$ than their
counterparts in the normal galaxies (squares), confirming that NRSF in these
LIRGs are indeed more active than in the normal galaxies. Additionally, most
NRSF regions in these LIRGs have SFR as high as the integrated ring values of
the normal galaxies, with 1--2\,dex smaller effective sizes, exhibiting
extremely high spatial concentration of SF activities.\\
\indent It is worth noting that for both the normal galaxies and the
LIRGs, regions measured at the highest physical resolutions (i.e. those in IC
342 and NGC 1614) have the highest $\Sigma_{\mathrm{SFR}}$, which demonstrates
the importance of high-resolution observations in accurate characterization of
these regions. Lower resolution observations would likely result in diluted (and
thus lower) measures of the intrinsic $\Sigma_\mathrm{SFR}$. Assuming the extreme
case where $\Sigma_\mathrm{SFR}$ is diluted by quiescent regions with no active
SF when measured within larger area (i.e. $\Sigma_\mathrm{SFR} \propto
1/R^2_e$, hatched in red in Figure \ref{fig:sigsfr_re}), NRSF regions
in NGC 1614 may appear to have similar $\Sigma_\mathrm{SFR}$ to resolved regions
in the normal galaxies even at 0\farcs3 resolution (100\,pc). The fact that
regions in NGC 1797 and NGC 7469 share similar $\Sigma_{\mathrm{SFR}}$ with
regions in IC 342 is likely a result of such dilution effect, given that
observations of these two galaxies have much lower physical resolutions.\\
\indent Furthermore, many regions in the LIRGs, especially in NGC 1614 are
unresolved by the beam (marked with arrows), which means that their sizes can be
even smaller, and their $\Sigma_\mathrm{SFR}$ can be even higher. Additionally,
the integrated ring values for these LIRGs all lie above the ranges spanned by
their NRSF regions, suggesting that active SF takes place throughout these
nuclear rings, not only in the NRSF regions that we characterized here. Indeed,
the sum of SFR in the NRSF regions only account for about 20 -- 50\% of the
total SFR of the rings. If higher resolution deep observations
were to be made available for these rings, we
would expect to detect more NRSF regions that are less
luminous or much smaller. \\
\indent In summary, the NRSF regions studied in our sample of LIRGs
intrinsically have higher SFR and $\Sigma_\mathrm{SFR}$ with sizes similar to or
smaller than their counterparts in the sample of normal galaxies. Observations
with consistent, high physical resolution are crucial for accurate
characterization of these extreme, compact NRSF regions.\\
\indent We note that within our sample, we do not find evidence associating AGN
activity with NRSF, as SFR or $\Sigma_{\mathrm{SFR}}$ in NRSF regions do not
appear consistently higher or lower in AGN hosts. Existing measurements of AGN
strength indicators in the IR and X-ray of the host galaxies
\citep{stierwalt13,dale06,grier11} also do not reveal any correlation with the
NRSF properties studied here. 

\subsection{NRSF in these LIRGs have SFR and $\Sigma_{\mathrm{SFR}}$ comparable
to luminous SF regions at high-$z$} 
\begin{figure}[tb!]
    \epsscale{1.2}
    \plotone{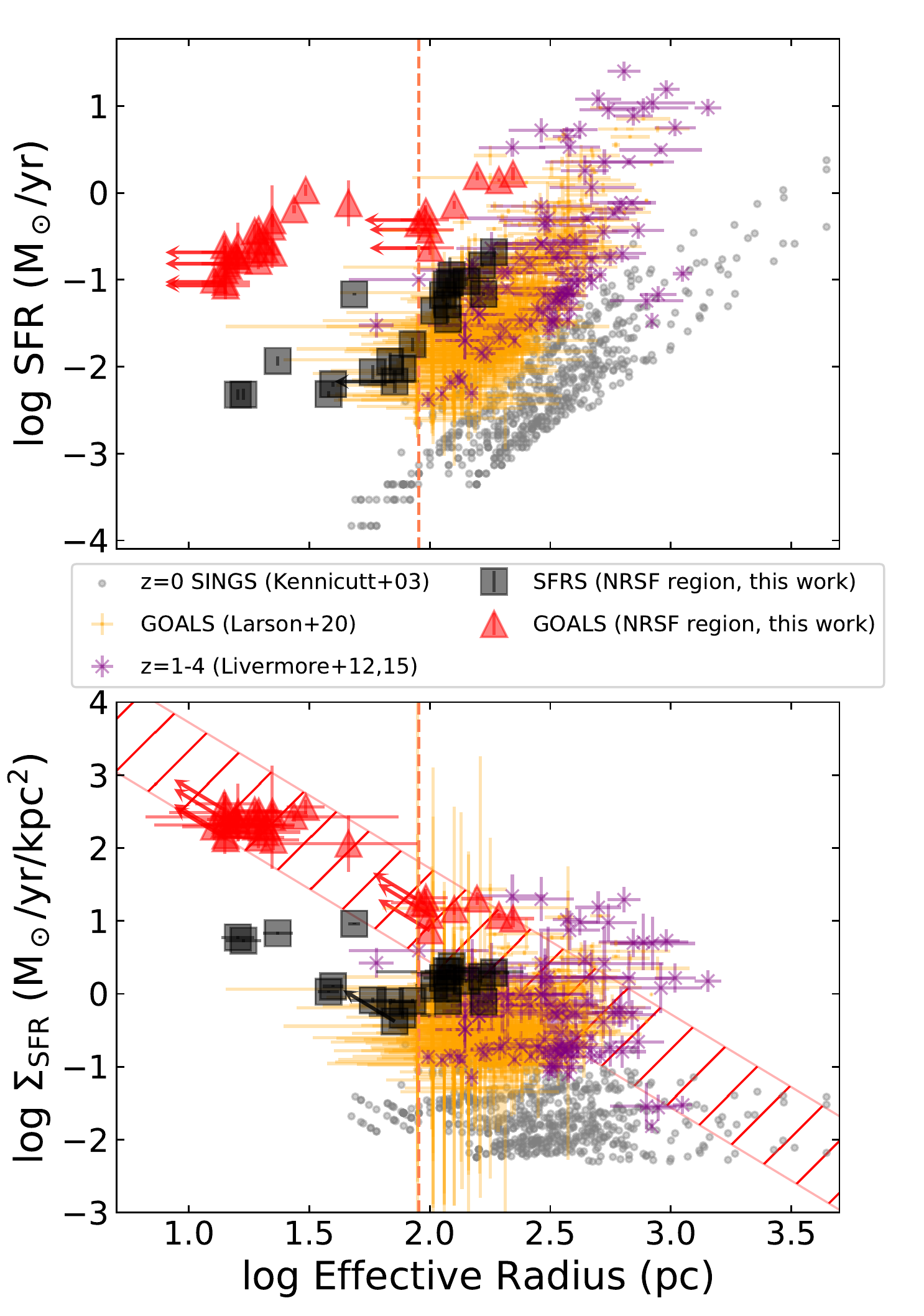} 
    \caption{Effective radius $R_e$ vs. SFR (\textit{top}) and
    $\Sigma_{\mathrm{SFR}}$ (\textit{bottom}) for SF regions in galaxies at
    different redshifts, reproduced from Figure 5 and 6 in \cite{larson20}.
    Triangles and squares are radio continuum measurements of individual NRSF
    regions in the LIRGs from the GOALS equatorial survey (red) and the normal
    galaxies from SFRS (black). Arrows indicate lower-limits in $R_e$ and
    upper-limits in $\Sigma_{\mathrm{SFR}}$ for unresolved regions. Purple ``x"
    represent H$\alpha$ measurements of SF regions in lensed galaxies at
    $z\simeq1-4$ \citep{livermore12,livermore15}. Orange ``+'' mark the
    Pa$\alpha$ and Pa$\beta$ measurements of 59 nuclei and 751 extra-nuclear SF
    regions in 48 local LIRGs from GOALS \citep{larson20}, with a resolution
    limit of 90\,pc (orange dashed vertical line). Grey dots are H$\alpha$
    measurements of SF regions in SINGS galaxies. \textbf{NRSF regions in the
    sample of local LIRGs have SFR and $\Sigma_{\mathrm{SFR}}$ comparable to or
    higher than luminous SF regions at high-$z$.}\label{fig:highz}}
\end{figure}

\indent Using high-resolution HST Pa$\alpha$
and Pa$\beta$ observations, \cite{larson20} measured the SFR and effective radii
of 751 extra-nuclear SF regions and 59 nuclei in 48 local LIRGs from GOALS. The
authors showed that SF in local LIRGs bridges the gap between the local and the
high-$z$ Universe, with a wide range of SFR overlapping with those found in
luminous SF clumps in $z = 1-4$ lensed galaxies. In Figure \ref{fig:highz}, we
reproduce Figure 5 and 6 from \cite{larson20}, overlaid with radio measurements
for the individual NRSF regions from this study. Note that the most
luminous regions in \cite{larson20} are the nuclei, most of which
have SFR $>$ 0.1\,M$_\odot$/yr and $\Sigma_\mathrm{SFR}$ $>$
0.2\,M$_\odot$/yr/kpc$^2$ with effective radii greater than 300\,pc. The
dashed vertical line in orange marks the resolution limit of 90\,pc for
measurements from \cite{larson20}. There are several conclusions that can be
drawn from this Figure:\\
\indent \textbullet \quad For the five normal galaxies from SFRS, the NRSF
regions (black squares) have higher SFR and $\Sigma_{\mathrm{SFR}}$ than the
ensemble of SINGS regions (grey dots), which are measured in the disk of normal
galaxies. Similarly, for the sample of LIRGs from the GOALS equatorial survey,
the NRSF regions (red triangles) have higher SFR and $\Sigma_{\mathrm{SFR}}$
than the ensemble of GOALS regions (orange ``+'') detected in the near-IR, over
90\% of which are extra-nuclear. These two results together suggest that NRSF
can be more extreme than extra-nuclear SF in the disk of the host galaxy,
supporting findings from \cite{linden19}.\\
\indent \textbullet \quad At similar measured sizes, NRSF regions in the sample
of normal galaxies overlap with the extra-nuclear GOALS regions in
\cite{larson20} and lower-luminosity lensed regions at high-$z$ (purple ``x'').
On the other hand, NRSF regions in the sample of LIRGs lie above the
extra-nuclear GOALS regions from \cite{larson20}, and their SFR and
$\Sigma_{\mathrm{SFR}}$ are comparable to many luminous high-$z$ regions. Note
that if we consider the same dilution analysis as shown in Figure
\ref{fig:sigsfr_re} (in red hatching), we would still find that the NRSF regions
in the LIRGs have $\Sigma_\mathrm{SFR}$ comparable to many high-$z$ SF regions
with lower resolution measurements. We will investigate whether this applies
more broadly to other nuclear SF regions in the GOALS equatorial survey in the
upcoming paper. Future surveys with the capability of detecting fainter and
smaller NRSF regions will allow a more comprehensive understanding of NRSF in
LIRGs. \\
\indent We also note that the different SFR tracers used for data
presented in Figure \ref{fig:highz} are sensitive to dust obscuration at
different levels, which can affect the above interpretation of our results.
While heavy non-uniform extinction in the nuclei of LIRGs can lead to
underestimation of SFR and $\Sigma_{\mathrm{SFR}}$ by 1 - 1.5\,dex even in the
near-IR \citep{u19}, over 90\% of the GOALS regions being compared here are
extra-nuclear and expected to be mildly extincted \citep{larson20}. Therefore we
do not expect extinction correction to change our conclusions. 

\subsection{Thermal fractions in the NRSF regions at 33 GHz}\label{dis:reg_fth} 
\begin{figure}[h!]
    \epsscale{1.2}
    \plotone{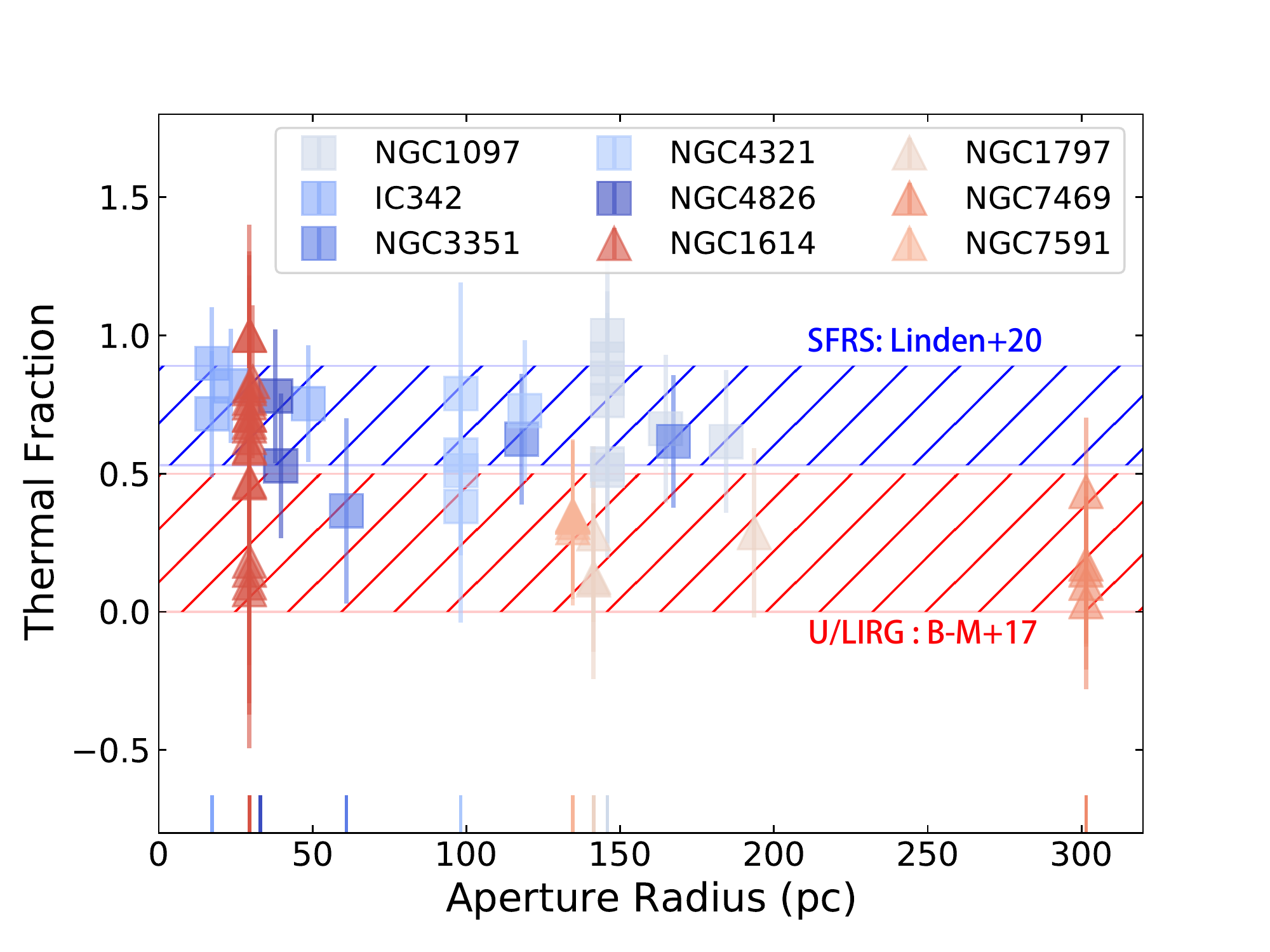} 
    \caption{Thermal fraction at 33 GHz with respect to the physical size of the
    aperture used for the measurements for the NRSF regions. Each region is
    color-coded by $L_\mathrm{IR}$ of its host galaxy, and squares and triangles
    represent normal galaxies and LIRGs separately. Short vertical lines
    on the horizontal axis indicate resolution limits of the spectral index
    measurements (see Section \ref{sec:astrodendro}). Blue hatched area marks
    the median values and median absolute deviation of 71\% and $\pm$ 18\%
    measured by \cite{linden20} on 100 $\sim$ 500\,pc scales for the nuclear
    regions ($r_G <$ 250\,pc) in the SFRS sample, and red hatched area represent
    the values for the nuclei in the most luminous local U/LIRGs ($\le $50\%)
    predicated by \cite{barcos17} at similar physical scales. \textbf{When
    measured at similar physical scales, NRSF regions in the sample of LIRGs
    have lower thermal fractions compared to regions in the sample of
    normal galaxies. NRSF regions in NGC 1614, which are measured at the
    smallest physical scales, span a wide range in thermal fraction.}
    \label{fig:fth_beam}}
\end{figure}

\indent In Figure \ref{fig:fth_beam}, we show the estimated thermal fractions at
33 GHz, $f_\mathrm{th}$, associated with the NRSF regions in the sample nuclear rings with
respect to the physical radii of the apertures within which these measurements were made
(see Section \ref{sec:sfregions} and \ref{sec:regionfth}).
For context, we also overlay the expected values from \cite{barcos17} for
luminous nuclei in U/LIRGs (hatched red), and the median values from
\cite{linden20} for the nuclear regions in SFRS (hatched blue), both
measured at sub-kpc scales. \\
\indent The median value of $f_\mathrm{th}$ associated with the NRSF
regions in the sample of normal galaxies is $\sim$ 69$\pm$19\%, in agreement
with the median value of $\sim$ 71\% for all nuclear regions (i.e. having
galactocentric radii $ r_G < 250$\,pc) in the full SFRS sample, reported in
\cite{linden20}. As discussed in \cite{linden20}, excess non-thermal emission is
present in the circum-nuclear SF regions in SFRS compared to the extra-nuclear
regions ($ r_G >= 1$\,kpc; median $f_\mathrm{th} \sim$ 90\%), likely due to
prolonged SF activities. As illustrated in Figure 7 of \cite{linden20} using
Starburst99 models, continuous SF for over 100\, Myr can decrease the thermal
fraction to $\sim$ 50\% due to accumulation of non-thermal emission from
supernovae, while instantaneous starburst can dramatically bring down thermal
fraction to much lower levels within 10\,Myr. Given that nuclear rings have
prolific episodic starbursts \citep[e.g.][]{buta2000,maoz01}, and can persist at
Gyr timescales \citep{knapen95, seo13}, the relatively low $f_\mathrm{th}$
observed in the NRSF regions in these normal galaxies may be driven by a
combination of continuous and ``bursty" SF.\\
\indent In Figure \ref{fig:fth_beam} we also see that at similar physical
scales, $f_\mathrm{th}$ can be even lower for NRSF regions in the sample of
LIRGs compared to those in the normal galaxies, except for NGC 1614, whose NRSF
regions span a wide range in $f_\mathrm{th}$. The median $f_\mathrm{th}$ in the
NRSF regions in the sample of LIRGs is $\sim$ 35$\pm$36\% including NGC
1614, and $\sim$ 29$\pm$9\% excluding NGC 1614, which are much lower
than the median of 69\% measured in the normal galaxies. This is in
agreement with findings from \cite{barcos15} and \cite{barcos17} that the
nuclear regions in U/LIRGs are mostly dominated by non-thermal emission. These
authors suggest that thermal emission in the nuclei of U/LIRGs may be
suppressed via the absorption of ionizing photons by dust. However, in our case,
the lower $f_\mathrm{th}$ can also be explained by beam dilution due to low
resolutions of the beam-matched images, i.e., measurements for most
regions in NGC 1797, NGC 7469 and NGC 7591 are at scales 2 - 5 times larger than
the sizes of the NRSF regions characterized using \textit{Astrodendro}. The
wide range of $f_\mathrm{th}$ observed in the nuclear ring of NGC 1614,
which is measured at physical scales smaller than 50\,pc, has also been
observed by \cite{herrero-illana14}. The authors conclude that this large
variation reflects the different ages of starbursts in the NRSF regions, with
regions of extremely young starbursts ($< 4\,$Myr) having thermal fractions
$\sim 100$\%, and regions of old starbursts ($> 8$\,Myr) having much lower
thermal fractions. This explanation is also consistent with Figure 7 in
\cite{linden20}. As demonstrated in previous Sections, the NRSF regions in the
sample of LIRGs are likely more compact than the ones in the normal galaxies.
Therefore, measurements at 100 -- 300\,pc scales in these LIRGs may average over
areas of young and old starbursts that have drastically different thermal
content. Additionally, as mentioned in Section \ref{sec:regionfth}, non-thermal
emission associated with supernovae can be more diffuse than thermal emission
\citep[e.g.][]{condon92}, hence low resolution measurements are more likely to
represent non-thermal emission. It is possible that, at high resolutions
(e.g. $<$ 100\,pc), we may also see very high $f_\mathrm{th}$ in some of the
NRSF regions in NGC 1797, NGC 7469 and NGC 7591, as observed in NGC 1614.
\subsection{Star formation relation}\label{sec:tdep} 
\begin{figure}[tb!]
     \epsscale{1.2}
     \plotone{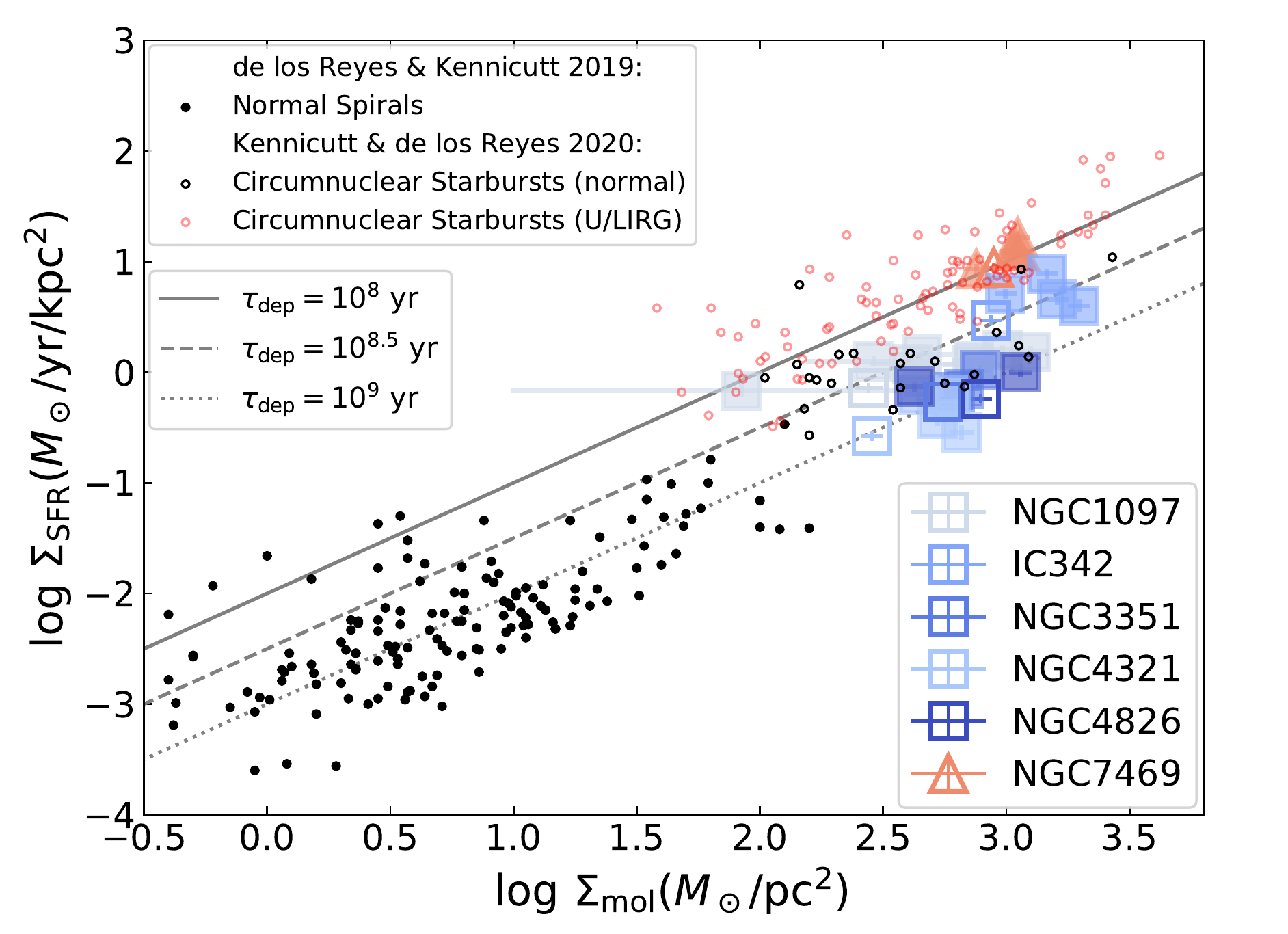}
     \caption{$\Sigma_\mathrm{mol}$ vs. $\Sigma_\mathrm{SFR}$ for the integrated
     nuclear rings (non-filled) and NRSF regions (filled), color-coded by the
     $L_\mathrm{IR}$ of the host galaxy. Also shown are global measurements for
     the normal spiral galaxies (black filled circles) from \cite{delosreyes19},
     and circum-nuclear starbursts in normal galaxies (black non-filled
     circles) and U/LIRGs (red non-filled circles) from \cite{kennicutt20},
     converted to match with the $\alpha_\mathrm{CO}$ we adopted. Solid, dashed,
     and dotted grey lines represent gas depletion time $\tau_\mathrm{dep}$ of
     $10^8$, $10^{8.5}$, and $10^{9}$ yr. \textbf{The estimated
     $\tau_\mathrm{dep}$ is shortest in the nuclear ring of NGC 7469, and has
     a large scatter among nuclear rings in the normal galaxies at similar
     $\Sigma_\mathrm{mol}$}.}\label{fig:k-s}
     \end{figure}

Global measurements of
$\Sigma_\mathrm{SFR}$ and $\Sigma_\mathrm{mol}$ of galaxies of various types
indicate the existence of a universal SF relation, i.e.
$\Sigma_\mathrm{SFR}=A\Sigma^{N}_\mathrm{mol}$, with $N$$\sim$1.4
\cite[e.g.][]{kennicutt98}. According to this relation, SF efficiency increases
(i.e. gas depletion time decreases) towards high $\Sigma_\mathrm{mol}$ for all
types of galaxies. However, several studies argue for a bimodal SF relation
which predicts constant star formation efficiency among galaxies of similar
types, with gas depletion times in normal spiral galaxies 4 - 10 times longer
than in U/LIRGs or high-$z$ sub-millimeter galaxies
\citep[e.g.][]{bigiel08,daddi10,genzel10,kennicutt20}. As we show in Figure
\ref{fig:k-s}, this bimodality is also present in our results for the nuclear
rings (non-filled) and the NRSF regions (filled) derived from extinction-free
measurements at sub-kpc scales (see Section \ref{sec:tdep_result}). Despite
having 1 - 2 dex higher $\Sigma_\mathrm{mol}$, the rings and the NRSF regions in
the normal galaxies (square symbols) have similar gas depletion times
($\tau_\mathrm{dep} \sim 1$\,Gyr) with normal spiral disks. The ring and NRSF
regions in NGC 7469 also show consistent $\tau_\mathrm{dep} (\sim 100$\,Myr)
with circumnuclear disk measurements for U/LIRGs, being up to an order of
magnitude shorter than rings and NRSF regions in the normal galaxies. This is
similar to values measured in other sub-kpc scale studies of U/LIRGs
\citep[e.g.][]{Xu15, p16}, but 4 - 6 times shorter than $\tau_\mathrm{dep}$
estimated from global measurements of GOALS galaxies \citep{herrero-illana19}.
In a resolved study of five U/LIRGs, \cite{wilson19} demonstrated that
$\tau_\mathrm{dep}$ decreases more rapidly at increasing $\Sigma_\mathrm{mol}$
above $\Sigma_\mathrm{mol} > 10^3$M$_\odot$/pc$^2$ in these extreme systems. In
an upcoming paper, we will present sub-kpc measurements for a larger sample of
nuclear SF regions in the GOALS equatorial survey to further explore the sub-kpc
SF relation in local U/LIRGs. \\
\indent The nuclear rings in IC 342 and NGC 1097 also have relatively
high SFE compared to spiral disks and other nuclear rings hosted in normal
galaxies, with $\tau_\mathrm{dep} \sim 0.4\,$Gyr. This central enhancement of
SFE has also been observed in other studies of IC 342 \citep{sage91,pan14} and
in surveys of normal galaxies \citep[e.g.][]{leroy13,utomo17}. Meanwhile, at
similar $\Sigma_\mathrm{SFR}$, NRSF regions in NGC 1097 span $\sim$ 1\,dex in
$\tau_\mathrm{dep}$. \cite{tabatabei18} discovered that this large scatter in
SFE is closely tied to local build-up of the magnetic field that support
molecular clouds against gravitational collapse. Overall, our results show that,
in these nuclear rings, at similar $\Sigma_\mathrm{mol}$, $\tau_\mathrm{dep}$ is
shorter in the LIRG NGC 7469 compared to in the normal galaxies, but it
varies among the normal galaxies as well, likely reflecting variation in local
SF conditions. Tentatively, this supports the idea of a multi-modal star
formation relation on sub-kpc scales. We note that adopting a normal galaxy
$\alpha_\mathrm{CO}$ for NGC 7469, or environmentally-dependent
$\alpha_\mathrm{CO}$\citep{narayanan12, sandstrom12}, can potentially produce a
more continuous SF relation among these nuclear rings, but more statistics
are needed to explore this.
\section{Summary}\label{sec:summary} In this paper we present analyses of
sub-kpc resolution VLA radio continuum observations of nine nuclear rings hosted
in four local LIRGs from the GOALS equatorial survey (NGC 1614, NGC
1797, NGC 7469, NGC 7591) and five nearby normal galaxies from the Star
Formation in Radio Survey (NGC 1097, IC 342, NGC 3351, NGC 4321, NGC 4826).
These two surveys map the brightest 3, 15 and 33 GHz radio continuum emission in
56 nearby normal galaxies and 68 local U/LIRGs at matched physical resolution,
and hence allow direct, extinction-free comparison of nuclear star formation
across different host environments. Using high-resolution maps of 33 or 15 GHz
continuum, we characterize the size, SFR and $\Sigma_\mathrm{SFR}$ of these nine
detected nuclear rings and 58 individual NRSF regions at $\sim 100$\, scales. We
summarize our main findings as follows: \\

\indent 1. The five nuclear rings in normal galaxies contribute 7 -
40\% of the total SFR of their host galaxies, with radii, SFR and
$\Sigma_\mathrm{SFR}$ in the range of 43 -- 599\,pc, 0.03 -- 2.0\,M$_\odot$
yr$^{-1}$ and 0.27 -- 2.90\,M$_\odot$ yr$^{-1}$ kpc$^{-2}$, respectively. By
comparison, the four nuclear rings in the LIRGs have much more dominant
contributions to the total star formation of their host galaxies, at 49 - 60\%,
with radii, SFR and $\Sigma_\mathrm{SFR}$ in the range of 121 -- 531\,pc, 6.1 --
29 \,M$_\odot$ yr$^{-1}$ and 6.0 -- 97\,M$_\odot$ yr$^{-1}$ kpc$^{-2}$,
respectively.\\

\indent 2. We identified a total of 58 individual NRSF regions using
\textit{Astrodendro}, 22 of which are hosted in the five normal galaxies, and 35
are in the LIRGs. NRSF regions in the normal galaxies have effective radii, SFR
and $\Sigma_{\mathrm{SFR}}$ in the range of 16 -- 184\,pc, 0.01 --
0.21\,M$_\odot$yr$^{-1}$ and 0.4 -- 9.2\,M$_\odot$yr$^{-1}$kpc$^{-2}$,
respectively. NRSF regions in the LIRGs have similar range of effective radii,
of 13 -- 221\,pc, but their SFR and $\Sigma_{\mathrm{SFR}}$ are an order of
magnitude higher, with a range of 0.08 -- 1.7 \,M$_\odot$yr$^{-1}$ and 7 --
402\,M$_\odot$yr$^{-1}$kpc$^{-2}$, respectively. Many of these NRSF regions in
the LIRGs are unresolved by our observations, so they may be more compact with
higher intrinsic $\Sigma_{\mathrm{SFR}}$. We also found that these NRSF
regions in the LIRGs have SFR and $\Sigma_\mathrm{SFR}$ as extreme as measured
in lensed high-$z$ SF galaxies from the literature. \\

\indent 3. The median ratio of thermal emission to the total 33 GHz
radio continuum emission (i.e. thermal fraction) associated with the NRSF regions is
69$\pm$19\% in the normal galaxies, and 35$\pm$36\% in the LIRGs, which is lower
than estimates for extra-nuclear SF regions, but consistent with results from
previous studies. The dominant presence of non-thermal emission in the LIRGs may
originate from suppression of thermal emission due to absorption of ionizing
photons by highly concentrated dust in HII regions. In our case, it is more
likely due to insufficient resolution of the measurements that results in the
inclusion of more diffuse non-thermal emission from cosmic rays accelerated by
supernovae. A wide range of thermal fractions were observed in
NGC 1614 at high resolution ($< 100$\,pc), which likely reflects different ages
of the starbursts along the nuclear ring.\\

\indent 4. For all five normal galaxies and one LIRG (NGC 7469), we use
available archival CO(J=1-0) data with comparable resolutions to our 33 GHz
observations to further study star formation efficiencies in these nuclear rings
at sub-kpc scales. The nuclear rings and NRSF regions in the normal galaxies
have gas depletion times $\tau_\mathrm{dep}\sim$ 1\,Gyr, about an order
magnitude longer than in the nuclear ring and NRSF regions of NGC 7469
($\tau_\mathrm{dep}\sim$ 100\,Myr), which is consistent with results from previous
studies on kpc and global scales. However, $\tau_\mathrm{dep}$ estimated for
rings and regions with similar $\Sigma_{\mathrm{mol}}$ have fair amount of
scatter, which may point to a multi-modal star formation relation on sub-kpc
scales. More statistics are needed to explore this. \\

\indent In this work we have demonstrated the ability to study embedded nuclear
ring star-forming regions on sub-kpc scales in local LIRGs using high frequency
radio continuum as an extinction-free tracer of star formation. This makes it
possible to directly compare star formation properties in the heavily obscured
hearts of local U/LIRGs with those at the center of normal galaxies. We also
show that to fully resolve and characterize these extremely compact NRSF regions
in the local LIRGs, observations at even higher resolutions and better
sensitivity are needed. Future facilities such as the ngVLA will greatly improve
our understanding of deeply embedded compact nuclear structures in these
systems.\\

\acknowledgments
We thank the anonymous reviewer for providing detailed and constructive feedback
that significantly improved this manuscript. We also thank M. de los Reyes for
sharing with us the datasets from \cite{kennicutt20}, and V. Casasola and S.
Ishizuki for making their data for NGC 4826 and IC 342 available online. Y.S.
thanks the staff at National Radio Astronomy Observatory for providing valuable
guidance on ALMA data reduction, and Alejandro Saravia, Eduardo Rodas-Quito and
William Meynardie for helpful discussions. A.S.E. and Y.S. were supported by NSF
grant AST 1816838. Y.S. and S.T.L. acknowledge support by the NRAO Grote Reber
Dissertation Fellowship. The National Radio Astronomy Observatory is a facility
of the National Science Foundation operated under cooperative agreement by
Associated Universities Inc. A.S.E. was also supported by the Taiwan, ROC,
Ministry of Science and Technology grant MoST 102-2119-M-001-MY3. H.I.
acknowledges support from JSPS KAKENHI Grant Number JP21H01129. V.U acknowledges
support from the NASA Astrophysics Data Analysis Program (ADAP) grant
80NSSC20K0450. This research has made use of the NASA/IPAC Extragalactic
Database (NED), which is funded by the National Aeronautics and Space
Administration and operated by the California Institute of Technology. This
research has also made use of NASA's Astrophysics Data System.\\

This paper makes use of the
following ALMA data: ADS/JAO.ALMA\#2012.1.00001.S, \\
ADS/JAO.ALMA\#2013.1.00885.S,\\
ADS/JAO.ALMA\#2015.1.00978.S,\\
 ADS/JAO.ALMA\#2016.1.00972.S,\\
ADS/JAO.ALMA\#2013.1.00218.S. \\

ALMA is a partnership of ESO (representing its
member states), NSF (USA) and NINS (Japan), together with NRC (Canada), MOST and
ASIAA (Taiwan), and KASI (Republic of Korea), in cooperation with the Republic
of Chile. The Joint ALMA Observatory is operated by ESO, AUI/NRAO and NAOJ.

\facilities{VLA(NRAO),
            ALMA(NRAO), 
            NASA/IPAC Extragalactic Database (NED),
            NASA's Astrophysics Data System}

\software{Ned Wright's Cosmology Calculator \citep{wright06},
          Astropy \citep{astropy:2013, astropy:2018},  
          Astrodendro (http://www.dendrograms.org/), 
          CASA \citep{mcmullin07}
          }
\newline
\newline
\bibliography{reference}{}
\bibliographystyle{aasjournal}
\appendix
\vspace{-0.5cm}
\section{Notes on Individual Galaxies}\label{sec:sample_detail}
Here we provide description of individual sources based on data used in this
work, as displayed in Figure \ref{fig:natres1}, \ref{fig:natres2},
\ref{fig:lightprofile} and \ref{fig:ka_co}. Where relevant, we include
descriptions from prior studies of each nuclear ring.\\

\indent \emph{NGC 1097}: In all three VLA bands, we clearly see a nearly
circular star-forming ring with diameter $(D)\sim17''$ (1.2\,kpc) made of
multiple bright SF knots, surrounding a luminous nucleus. Emission from the
nucleus and the ring are well separated in the azimuthally averaged light
profile at 33 GHz, which allows us to characterize the spatial extent of the
ring. A nuclear spiral inside the ring, which is transporting gas into the
nucleus through the ring, has been revealed with multi-wavelength observations
by \cite{prieto19}. Our radio images also show three faint streamer-like
structures that connect the central nucleus to the ring, with the brightest
streamer on the west of the nucleus extending few arcseconds beyond the ring.
This extension has a bright counterpart in the ALMA CO(J=1-0) data, which
coincides with the contact point of a kpc gas streamer feeding in ring, as
discovered by \cite{prieto19}. The host galaxy ($cz = 1270$\,km/s) is interacting with at least
one dwarf companion \citep[1097A: $cz = 1368$\,km/s;][]{bowen16}. \\

\indent \emph{IC 342}: A small asymmetric nuclear star-forming ring
($D\sim 6'',160$\,pc) made of at least four distinct SF knots is detected at all three VLA
bands. At 33 GHz, diffuse emission from the ring
covers the central region, as seen in the azimuthally-averaged light
profile. No bright nuclear emission is detected at any VLA Band, and
the ring morphology has previously been confirmed in near-IR and CO observations
\citep{boker97,schinnerer03}. Therefore we do not assign an inner radius for
this nuclear ring when calculating its SFR to account for the diffuse emission,
but the area of the synthesized beam is subtracted from the area defined by the
outer radius when estimating $\Sigma_\mathrm{SFR}$ to account for the central
cavity. In a previous molecular gas study, \cite{ishizuki90} suggest that the nuclear ring
outlines the ends of a pair of molecular ridges, which may have formed due to
shock-waves from a bar-like gravitational potential.\\

\indent \emph{NGC 3351}: An elliptical nuclear ring ($D\sim13'',600$\,pc) with
an inclination angle of $\sim 59^\circ$ is clearly present in images at all
three VLA bands, with two bright SF knots lying on the north and south tips of
the ring separately, accompanied by smaller, fainter SF knots on the east and
west sides. The azimuthally-averaged light profile at 33 GHz reveals a faint
nuclear component, which outlines the inner radius of the ring. This component
is most visible at 3 GHz. A comparison of the radio-detected ring with archival
Spitzer IRAC 3.6$\mu$m image of the galaxy reveals a bar-like stellar structure
connecting the brightest radio ``hotspots'' along the north-south direction. Low
resolution ($\sim 7\arcsec$) molecular observations have suggested the presence
of a molecular nuclear bar \citep{devereux92}, which is absent from the ALMA
CO(J=1-0) observation. \\

\indent \emph{NGC 4321}: The nuclear ring ($D\sim14'',1$\,kpc) was detected at
all three bands, along with the central LINER nucleus. The ring appears fairly
clumpy, with three bright knots making up the east half of the ring and
relatively faint diffuse emission on the west half. At 3 GHz more diffuse
emission is detected, and the ring reveals itself to be part of a tightly wound
spiral structure, which is also evident in the ALMA CO(J=1-0) data. The ends of
the molecular spiral arms correspond to the location of bright SF knots, which
connects the ring to a nuclear bar that is prominent in both archival Spitzer
IRAC 3.6 $\mu$m image, and CO data \citep{sakamoto95}. Numeric simulations
predict that such nuclear bars exert strong gravitational torques on molecular
gas to effectively feed the SMBHs and nuclear/circum-nuclear starbursts
\citep{wada98}. \\

\indent \emph{NGC 4826}: The small nuclear ring ($D\sim5'',120$\,pc) is
most apparent at 15 GHz, where we see five distinct knots lying close together
to form a nearly circular structure. The brightest knot is the off-center LINER
nucleus \citep{garcia03}. At 33 and 3 GHz, the emission from the two fainter
knots is blended into the brighter knot in between them. The bright nuclear ring
is also closely surrounded by a series of much fainter SF regions, which
show up in the azimuthally-averaged light profile at $\sim10''$
(250\,pc) away from the center (we exclude them from our analyses). At 33 GHz,
diffuse emission from the ring covers the central region, as seen in the light
profile. No bright nuclear emission is detected at any VLA Band, and the ring
morphology has previously been confirmed in CO observation \citep{garcia03}.
Therefore we do not assign an inner radius for this nuclear ring when
calculating its SFR to account for the diffuse emission, but the area of the
synthesized beam is subtracted from the area defined by $R_\mathrm{out}$ when
estimating $\Sigma_\mathrm{SFR}$ to account for the central cavity.\\

\indent \emph{NGC 1614}: A clumpy, almost circular nuclear star-forming ring
($D\sim1.5'',400$\,pc), made of knots with various
sizes and brightness, is detected at 33 and 15GHz. One faint elongated knot on the west extends beyond the
ring by $\sim 0.5''$ (150\,pc). A faint nucleus surrounded by the ring is also
detected at 33 GHz, which outlines the inner radius of the ring. At 3 GHz, the
ring is unresolved. Prominant dust lanes have been observed to be connected to
the northern tip of the nuclear ring where molecular gas is potentially streaming
into the ring and fueling the starbursts \citep{olsson10,konig13}. \\

\indent \emph{NGC 1797}: The ring structure is resolved for the very first time
with the GOALS equatorial survey. At 33 GHz, we clearly see three bright SF
knots connected by diffuse emission, forming an elliptical ring
($D\sim2'',800$\,pc) with an inclination angle of $\sim 45^\circ$. At 15 GHz,
the brightest region on the east half of the ring is further resolved into three
smaller distinct SF knots, with two brighter ones on the north connected to each
other. The ring morphology at 3 GHz follows that at 33 GHz, but the diffuse
emission connecting the east and west half the ring becomes more prominent. No
bright nuclear emission is detected at any VLA Band, therefore we do not assign
an inner radius for this nuclear ring when calculating its SFR to account for
the diffuse emission, but the area of the synthesized beam is subtracted from
the area defined by $R_\mathrm{out}$ when estimating $\Sigma_\mathrm{SFR}$ to
account for the central cavity.\\

\indent \emph{NGC 7469}: The ring structure ($D\sim3'',1000$\,pc)
containing five SF knots is clearly detected at 33GHz, surrounding a much
brighter Seyfert 1 nucleus, which outlines the inner radius of the ring. The
ring appears more spiral-like at 3 GHz, where the northern and southern
components are much more pronounced and extending beyond the ring. A similar
morphology is also seen in ALMA CO(J=1-0) data. Due to low angular
resolution, the ring appears as faint diffuse emission surrounding the bright
nucleus at 15 GHz. \cite{mazzarella94} observed the nuclear ring in the near-IR,
and found the brightest SF ``hotspots'' coincide with the ends of a nuclear
stellar bar revealed in K-band continuum, which may be transporting gas from the
ring to the luminous Seyfert 1 nucleus \citep[e.g.][]{wada98}. \\

\indent \emph{NGC 7591}: The elliptical ring ($D\sim1'',300$\,pc) has an
inclination angle of $\sim 62^\circ$, and is detected at all three bands but was
only resolved with at least $3\sigma_\mathrm{rms}$ at 15 GHz, with the southern
part of the ring brighter than the rest. At 3 and 33 GHz, the ring becomes
completely unresolved. The nuclear ring has also been observed in the near-IR
Paschen observations by \cite{larson20}, but NRSF regions in the rings are
resolved for the first time with the GOALS equatorial survey.

\section{Integrated measurements for highly elliptical rings}\label{sec:integrated_method}
For all but three galaxies (NGC 1797, NGC 3351 and NGC 7591), we measured the
azimuthally-averaged light profiles of each nuclear ring by computing the averaged brightness 
per pixel on a series of 1 pixel wide concentric circles overlaid on
top of 33GHz images, with their centers aligned with the central coordinates of
the host galaxy, with minor adjustments to visually match the ring center. Due to
relatively high ellipticities of the nuclear rings in NGC 1797, NGC 3351 and NGC
7591, we used the following procedures to more accurately depict their light
profiles: After each image is masked to only preserve emission with SNR$ > 5$ ,
the coordinates of the unmasked pixels were extracted and then fitted with a 2D
ellipse model using the least square fitting method suggested in
\citet{fitzgibbon96}. Note that 15 GHz image was used for NGC 7591 instead due
to low resolution of the available 33 GHz image. The model describes a generic
quadratic curve:
\begin{equation} \label{eqn:ellipse}
    ax^2+2bxy+cy^2+2dx+2fy+g=0
\end{equation}
where the fitted ellipse's center coordinate $(x_0, y_0)$, semi-major and
semi-minor axes lengths $(a', b')$ and the counterclockwise angle of rotation
from the x-axis to the major axis of the ellipse could be calculated as follows:
\begin{align*}
    x_0 &= \frac{cd-bf}{b^2-ac}\\
    y_0 &= \frac{af-bd}{b^2-ac}\\
    a'&=\sqrt{\frac{2(af^2+cd^2+gb^2-2bdf-acg)}{(b^2-ac)[\sqrt{(a-c)^2+4b^2}-(a+c)]}}\\
    b'&=\sqrt{\frac{2(af^2+cd^2+gb^2-2bdf-acg)}{(b^2-ac)[-\sqrt{(a-c)^2+4b^2}-(a+c)]}}\\
    \phi&=\begin{cases}
      0, & \text{for}\ b=0 \ \text{and}\ a<c \\
      \frac{\pi}{2}, & \text{for}\ b=0 \ \text{and}\ a>c \\
      \frac{1}{2}\cot^{-1}{(\frac{a-c}{2b})}, & \text{for}\ b\neq0 \ \text{and}\ a<c \\
      \frac{\pi}{2}+\frac{1}{2}\cot^{-1}{(\frac{a-c}{2b})}, & \text{for}\ b\neq0 \ \text{and}\ a>c
    \end{cases}
\end{align*}
The above method works under the condition that $\begin{vmatrix} a & b & d \\ b
& c & f \\ d & f & g \end{vmatrix} \neq 0$, $\begin{vmatrix} a &b \\ b& c
\end{vmatrix}> 0$ and $(a+c) < 0$, which makes Eq.\ref{eqn:ellipse} the general
expression for 2D ellipses. The caveat is that this method does not work well
for circular fits, therefore we only adopt it to extrapolate the
properties of the highly elliptical rings in NGC 1797, NGC 3351 and NGC 7591. Based on the
ellipticity estimated from the model fitting result, we then produced a series
of concentric 1-pixel-wide elliptical annuli to calculate the adjusted
azimuthally-averaged brightness of the ring.\\
\end{document}